\documentclass[twocolumn, twocolappendix]{aastex701}
\usepackage{tikz}
\usepackage{amsmath}

\usepackage[normalem]{ulem} 

\newcommand{\delone}[1]{}

\newcommand{\revone}[1]{#1}

\begin{document}

\title{Multi-Wavelength Signatures of a Giant Cometary Radio Halo in MACSJ0417–1154}

\author[0009-0002-0373-570X]{Ramananda Santra}
\affiliation{National Centre for Radio Astrophysics, Tata Institute of Fundamental Research, Post Bag 3, Ganeshkhind, Pune 411007, Maharashtra, India}

\affiliation{International Centre for Theoretical Sciences, Tata Institute of Fundamental Research, Bangalore 560089, India}
\email[show]{ramananda1999@gmail.com}

\author[0009-0001-3048-0020]{Marco Balboni}
\affiliation{Dipartimento di Fisica e Astronomia (DIFA), Università di Bologna, via Gobetti 93/2, 40129 Bologna, Italy}

\affiliation{Istituto Nazionale di Astrofisica – Istituto di Astrofisica Spaziale e Fisica Cosmica (IASF), Via A. Corti 12, 20133 Milano, Italy}
\email{marco.balboni@inaf.it}

\author[0000-0003-1449-3718]{Ruta Kale}
\affiliation{National Centre for Radio Astrophysics, Tata Institute of Fundamental Research, Post Bag 3, Ganeshkhind, Pune 411007, Maharashtra, India}  
\email{ruta@ncra.tifr.res.in}

\begin{abstract}

Galaxy clusters hosting diffuse non-thermal radio emission offer direct insight into plasma processes of the intracluster medium (ICM). We present the first multi-frequency study of the radio halo in MACSJ0417 (z = 0.445) using uGMRT (300--850 MHz), MeerKAT (900--1670 MHz), and archival \textit{XMM-Newton} data. The halo extends to $\sim$1.75 Mpc at 400 MHz, while two candidate relics (R1 and R2) are detected at 2.9 Mpc. The integrated spectra follow single power-laws with spectral indices $\alpha \simeq -1.3$ for the halo and $\alpha \simeq -1.6$ for the relics. Sensitive uGMRT imaging reveals a radio surface brightness edge $\sim$43$''$ SE of the cluster centre, which coincides with an X-ray discontinuity. Resolved spectral maps (400--1280 MHz) show significant fluctuations and a clear radial steepening of the spectral index. X-ray analysis reveals an elongated SE–NW morphology and high temperature regions ($\sim$11 keV) along this axis. A strong radio and X-ray surface brightness correlation is found (correlation coefficient $\sim$ 0.85), with the correlation slope evolving from sublinear at 400 MHz to linear at 1280 MHz. These results, together with the spectral properties, support the turbulent re-acceleration model and point to inhomogeneous ICM conditions. The pure hadronic model is excluded owing to unrealistic energy requirements for cosmic-ray protons. We propose that MACSJ0417 is undergoing a minor off-axis dissociative merger (mass ratio $\sim$6:1) along the SE–NW axis, which has preserved its cool core while driving turbulence that powers the giant radio halo.

\end{abstract}


\keywords{Galaxy clusters(584)-Large-scale structure of the universe(902)-Intracluster medium(858)- Non-thermal radiation sources(1119)-Extragalactic radio sources(508)}


\section{Introduction} \label{sec:intro}

Galaxy clusters and their dark matter halos grow through accretion of sub-clusters and diffuse gas along cosmic filaments \citep[e.g.,][]{springel05,nelson24}. Major mergers, a key step in this process, inject energy of $\sim 10^{64}$ erg into the intracluster medium (ICM) by driving shocks and turbulence \citep[e.g.,][]{forman1982,sarazin1986}, which heats the ICM, accelerates cosmic rays (CRs), and amplifies magnetic fields. These processes produce diffuse synchrotron radio emission, known as radio halos and relics \citep[reviews:][]{feretti12,vanweeren19}. \revone{Despite diffuse radio emission having been studied for several decades and now being detected in hundreds of galaxy clusters} \citep[e.g.,][]{kale13, kale15, botteon22a, knowles22}, the micro-physics of energy transfer across different scales remains unclear \citep[for reviews][]{bode19, brunetti14}.

Radio relics are diffuse synchrotron sources typically found in the peripheral regions of merging galaxy clusters \citep[e.g.,][]{vanweeren19}. They show elongated or arc-like morphologies, physical extents from a few hundred kiloparsecs to nearly 2 Mpc, and significant linear polarisation, sometimes reaching levels of $\sim 30$\% \citep[e.g.,][]{vweeren10, 2017A&A...600A..18K, digennaro18, degasperin22, Pal2025}. Their peripheral locations, ordered magnetic fields, and frequent association with X-ray surface-brightness or temperature discontinuities support the interpretation that relics trace merger-driven shock waves in the intracluster medium \citep[e.g.,][]{ensslin98, hoeft07, rajpurohit18, 2020A&A...634A..64B, nuza24}.

Radio halos are centrally located sources, extend to Mpc scales (typically 1$–$2 Mpc or more), co-spatial with the ICM thermal emission, and show steep spectra\footnote{S$_{\nu} \propto \nu^{\alpha}$, where S$_{\nu}$ is the flux density at frequency $\nu$ and $\alpha$ the spectral index} with $\alpha < -1$. Statistical studies reveal strong links between radio halo properties and cluster dynamics \citep[e.g.,][]{cassano10,cassano13,kale13,kale15,cuciti21a}, supporting re-acceleration by merger-driven turbulence for the halo occurrence \citep[e.g.,][]{brunetti01,petrosian2001nonthermal,brunetti07,brunetti16}. However, the origin of seed relativistic electrons behind the radio halo origin remains uncertain, \revone{as in the current models, merger-driven turbulence is thought to re-accelerate a pre-existing population of mildly relativistic electrons to energies sufficient to produce the observed synchrotron emission} \citep[e.g.,][]{vazza21,vazza24}. \revone{The} hadronic model proposes cosmic-ray protons generate secondary electrons via collisions with thermal protons, producing synchrotron emission \citep[e.g.,][]{dennison80,blasi99,dolag00,brunetti17}. Yet, the absence of $\gamma$-rays from clusters with radio halos \citep{brunetti12,ackermann2012,ackermann2014,adam21} and the discovery of ultra-steep spectrum halos \citep{brunetti08,dallacasa2009,macario10,venturi13,rajpurohit23,santra24a,pasini24} suggest the contributions from the secondary electrons are subdominant. Still, scenarios where secondaries are re-accelerated by turbulence remain feasible \citep[e.g.,][]{brunetti05,brunetti11,nishiwaki24}.

Radio halos at intermediate redshifts (0.3 $\leq z \leq$ 0.6) provide a key window into the evolution of diffuse synchrotron emission in galaxy clusters \citep[e.g.,][]{martinez10}. While radio halos in nearby clusters ($z < 0.3$) are relatively well studied \citep{vanweeren19}, \revone{the properties of halos at $z > 0.3$ remain less well understood}. Models predict many halos at $0.3-0.6$ \citep{cassano2004,cassano06}, indicating that particle acceleration and magnetic field amplification were already active when clusters were less evolved than locally. This redshift range bridges the nearby universe and the high-$z$ regime ($z > 0.6$), where clusters are still assembling and rapid acceleration is required to offset Inverse Compton losses \citep{digennaro21,sikhosana2025,kale25, santr2026_ska}. \revone{Studying halos in this redshift regime} thus probes the non-thermal ICM energy budget and constrains models of particle acceleration and magnetic field growth across cosmic time \citep[e.g.,][]{xu2010,rappaz24}.

In this work, we present a multi-frequency radio and X-ray analysis of the galaxy cluster MACSJ0417$-$1154 (hereafter MACSJ0417), a massive merging system at $z=0.445$ \citep[e.g.,][]{2000A&AS..144..247C}. It is a hot cluster with T$_{500} \sim 11$ keV, an X-ray luminosity of (29.1 $\pm$ 0.5) $\times$ 10$^{44}$ erg~s$^{-1}$ in the 0.3$-$5 keV band, \delone{and a total mass of (22.1 $\pm$ 3.9) $\times$ 10$^{14}$ M$_{\odot}$ within r$_{500}$} \citep{2011A&A...534A.109P}. \textit{Chandra} observations show a comet-like diffuse morphology with a compact, bright core \citep{2012MNRAS.420.2120M}. X-ray data suggested a surface brightness jump, later supported by \textit{Chandra}, which revealed two cold fronts in the southeast (SE) and northwest (NW) direction from the cluster centre \revone{\citep{Botteon2018}}. \revone{Sunyaev--Zel'dovich (SZ) observations yield a mass of M$_{500}=(12.25\pm0.52)\times10^{14}$ M$_{\odot}$ \citep{2017MNRAS.464.2752P}}, and the weak-lensing analysis modelled the system as a main cluster and subcluster with a $\sim$6:1 mass ratio, yielding M$_{200}=(11.5 \pm 3.5)\times 10^{14}$ M$_{\odot}$ for the main cluster and M$_{200}=(1.96 \pm 0.95)\times 10^{14}$ M$_{\odot}$ for the subcluster, leading to a total mass of (13.8 $\pm 2.6)\times 10^{14}$ M$_{\odot}$ \citep{pandge19}.

This cluster hosts a $\sim$1 Mpc halo that was first discovered by \citet{2011JApA...32..529D} and later examined with GMRT (235, 610 MHz) and JVLA (1575 MHz) data by \citet{2017MNRAS.464.2752P}. \citet{2019JApA...40...17S} studied the cluster from 76 MHz to 18 GHz—the broadest spectrum for any halo—reporting a single power law with spectral index $-1.5$, without any high-frequency steepening. MeerKAT L-band (900–1670 MHz) observations from the MGCLS (MeerKAT Galaxy Cluster Legacy Survey\footnote{\url{https://mgcls.sarao.ac.za/}}) further revealed complex diffuse structures and confirmed large-scale non-thermal emission \citep{knowles22, kolokythas25}. However, spectral studies below 1 GHz remain limited by poor \textit{uv}-coverage and sensitivity, hindering full insight into the origin and evolution of the halo. Combined X-ray and lensing evidence classified MACSJ0417 as a dissociative merger, where gas is spatially offset from dark matter \citep{pandge19}. Such systems are unique laboratories for probing merger-driven turbulence and their role in the non-thermal emission.

This paper is organised as follows. The uGMRT, MeerKAT, and \textit{XMM-Newton} observations and data analysis procedures are explained in Section~\ref{sec:data analysis}. In Section~\ref{radio_analysis}, we show the uGMRT and MeerKAT continuum images at different resolutions and discuss newly discovered features from our study. The results obtained from radio spectral analysis, X-ray analysis, and radio vs X-ray correlations are described in Section~\ref{radio_analysis}$-$\ref{rad-xray-corr}. Then we summarised our results and findings in Section~\ref{summary}. Throughout this paper, we have adopted a flat $\Lambda$CDM cosmology with H$_{0}$ = 70 km s$^{-1}$, $\Omega$ $_{m}$ = 0.3 , $\Omega_{\Lambda} =0.7$. At the redshift of MACSJ0417, 1$''$ corresponds to a linear scale of 5.7 kpc.

\begin{table}
  \centering
  \caption{Summary of radio observations.}
  \begin{tabular}{@{}lccc@{}}
    \hline
      &  Band 3  & Band 4 & MeerKAT \\
    \hline
Frequency (MHz) & 300-500 & 550-750 & 900-1670   \\ 
 
 Bandwidth (MHz) & 200/156 & 200/177 & 856/770 \\
 
 Integration time (Hr.) & 10  & 10 & 8  \\

 Largest angular scale & 1920$''$  & 1020$''$ & 1440$''$ \\

 Shortest baseline ($\lambda$) & 150  & 200 & 86  \\
   \hline
  \end{tabular}
  \label{obs_tab}
 \tablecomments{The bandwidth is provided as the total/effective}
\end{table}

\begin{table*}
  \centering
  \caption{Properties of the radio images.}
  \begin{tabular}{@{}ccccccc@{}}
    \hline\hline
    & Name & Beam Size ($''$, $^{\circ}$) & Robust & \textit{uv} range & \textit{uv}taper ($''$) & map rms ($\mu$Jybeam$^{-1}$)  \\
    \hline
    
    uGMRT band 3 & IMG1 & 8.5$''$ $\times$ 6.5$''$, 60.5  & 0.5 & None & None & 27.0  \\ 

 & IMG2 & 15.0$''$ $\times$ 15.0$''$, 0.0 & 0.5 & \textgreater 0.2k$\lambda$& 10$''$ & 45.0   \\

   & IMG3 & 30.0$''$ $\times$ 30.0$''$, 0.0  & 0.5 & \textgreater 0.2k$\lambda$ & 25$''$ & 65.0 \\

 \hline

uGMRT band 4 & IMG4 & 5.0$''$ $\times$ 4.0$''$, 60.9  & 0.5 & None & None& 9.0    \\ 

 & IMG5 & 15.0$''$ $\times$ 15.0$''$, 0.0 & 0.5 & \textgreater 0.2k$\lambda$& 10$''$ & 28.7  \\

  & IMG6 & 30.0$''$ $\times$ 30.0$''$, 0.0  & 0.5 & \textgreater0.2 k$\lambda$& 25$''$ & 39.0  \\

\hline

MeerKAT & IMG7 & 8.0$''$ $\times$ 7.0$''$, -3.2  & 0 & None & None& 5.0  \\ 

 & IMG8 & 15.0$''$ $\times$ 15.0$''$, 0.0  & 0 & None & None& 35.0    \\

& IMG9 & 30.0$''$ $\times$ 30.0$''$, 0.0  & 0.5 & \textgreater0.2 k$\lambda$& 25$''$ & 10.0  \\
 
\hline
\end{tabular}

\label{img_summary}
\end{table*}

\section{Observations and data analysis} \label{sec:data analysis}

MACSJ0417 was observed with the upgraded GMRT (observation code: 45\_054), spanning a frequency coverage of $300-850$ MHz. We have used the uGMRT online RFI (Radio Frequency Interference) filtering \citep{buch19,buch23}, to mitigate the broadband RFI during the observations. MeerKAT L-band images are publicly available as part of the MGCLS survey; however, we have reprocessed the MeerKAT data. The observational parameters are summarised in Table~\ref{obs_tab}.

\subsection{uGMRT} \label{ugmrt_data_analysis}

We processed the uGMRT data using the \texttt{CASA} software with the aid of the \texttt{CAPTURE}\footnote{\url{https://github.com/ruta-k/CAPTURE-CASA6}} pipeline \citep{2021ExA....51...95K}, a continuum pipeline specifically designed for the reduction of GMRT continuum observations. Following initial flagging, the flux density of the primary calibrator was set according to the \texttt{Perley$-$Butler} 2017 flux scale \citep{2017ApJS..230....7P}. Standard calibration routines (bandpass and complex antenna gains) were performed in \texttt{CASA}, along with automated flagging, after which the calibration solutions were applied to the target field. The calibrated target visibilities were then split and subjected to additional automated flagging. Special attention was given to mitigating narrow-band RFI on short baselines at both frequencies using \texttt{AOFlagger} \citep{offringa12}. Imaging of the target field was carried out with \texttt{WSClean} \citep{offringa14}, using wide-field imaging techniques such as W-stacking to account for non-coplanar baselines. \revone{Several rounds of phase-only (five) self-calibration followed by three rounds of phase-and-amplitude self-calibration were performed to improve the calibration solutions and image dynamic range. For each self-calibration cycle, sky models were generated using \texttt{WSClean} and the resulting calibration solutions were derived and applied within \texttt{CASA}. This procedure was repeated independently for both observing bands.} For further details on the uGMRT data reduction procedure, refer to \citet{2021ExA....51...95K,2022MNRAS.514.5969K}.

\subsection{MeerKAT}\label{meerkat_data_analysis}

The data reduction consisted of two main steps: an initial calibration solution provided by the SARAO Science Data Processor (SDP), and subsequently performing additional rounds of self-calibration using the \texttt{facetselfcal} framework \citep{vanweeren21}, with some steps followed from \citet{Bott_24}. Specifically, we began by applying the default SDP calibration to our data using the \texttt{mvftoms.py} script from the \texttt{katdal} package. This step produced a calibrated measurement set, corrected for delay, bandpass, and gain effects. The data were then compressed using \texttt{Dysco} \citep{offringa16} and averaged by a factor of two in both time and frequency. Self-calibration was carried out with the \texttt{facetselfcal} pipeline \citep{vanweeren21}, which employs \texttt{DP3} \citep{vandipen18, dijkema23} for calibration and \texttt{WSClean} for imaging. Typically, we performed four rounds of self-calibration on the full MeerKAT field of view (FoV): two phase-only and two amplitude-and-phase calibration iterations. To streamline further processing, we then adopted the extraction and self-calibration strategy described by \citet{vanweeren21}. We refer to \cite{balboni25} for further details on the data analysis procedure.

\subsubsection{Discrete source subtractions \& flux density errors}\label{fluxcal_errors}

Measuring the flux of diffuse emission is challenging due to the presence of compact sources within the extended emission. To remove them, we modelled discrete sources using a \textit{uv} cut of $>$4k$\lambda$ ($\sim$50$''$), Fourier transformed the high-resolution model image, and subtracted it from the visibilities with \texttt{uvsub} in \texttt{CASA}. The source-subtracted residual data were imaged with \texttt{wsclean}, applying a \textit{uv}-cut of $<$10k$\lambda$, \texttt{robust=0.5}, and a \texttt{multiscale} deconvolver to recover extended emission. The error corresponding to the point source subtraction is estimated following the procedure where we compared the halo flux density derived after \textit{uv}-plane model subtraction with that obtained by
algebraically subtracting the flux densities of compact sources from the total (halo + sources). From this comparison, we estimate the subtraction error to be $\sim 8.5\%$ for uGMRT.

Flux density uncertainties were assumed to be 10\% at uGMRT frequencies \citep{chandra&kanekar17} and 5\% at MeerKAT frequencies\footnote{\url{https://archive-gw-1.kat.ac.za/public/meerkat/Modelling-of-MeerKAT-L-and-UHF-band-calibrators.pdf}} \citep{balboni25} 
The total uncertainty ($\Delta S$) is:
\begin{equation}
\Delta S = \sqrt{(f \cdot S)^{2} + N_{\rm beam} \cdot (\sigma_{\rm rms})^{2} + (\sigma_{\rm sub}\cdot S)^{2}},
\label{eq-flux-err}
\end{equation}
where $S$ is the flux density, $f$ is the absolute flux density calibration uncertainty, $N_{\rm beam}$ the number of beams, $\sigma_{\rm rms}$ the rms noise, and $\sigma_{\rm sub}$ is the discrete source subtraction error.

\begin{table}
\centering
\caption{Effective exposure time of each XMM-Newton -EPIC observation\label{xmm_exposures_tab}}
\begin{tabular}{lccc}
\hline\hline
Instruments & EMOS1 & EMOS2 & EPN \\ 
\hline
Med. live time (ks) & 65.9 & 65.9 & 63.4 \\ 
Med. on time (ks) & 42.3 & 41.5 & 44.6 \\ 
Time selection (\%) & 75.6 & 78.1 & 58.0 \\ 
\hline
\end{tabular}
\end{table}

\begin{figure*}
    \centering
    \includegraphics[width =\textwidth]{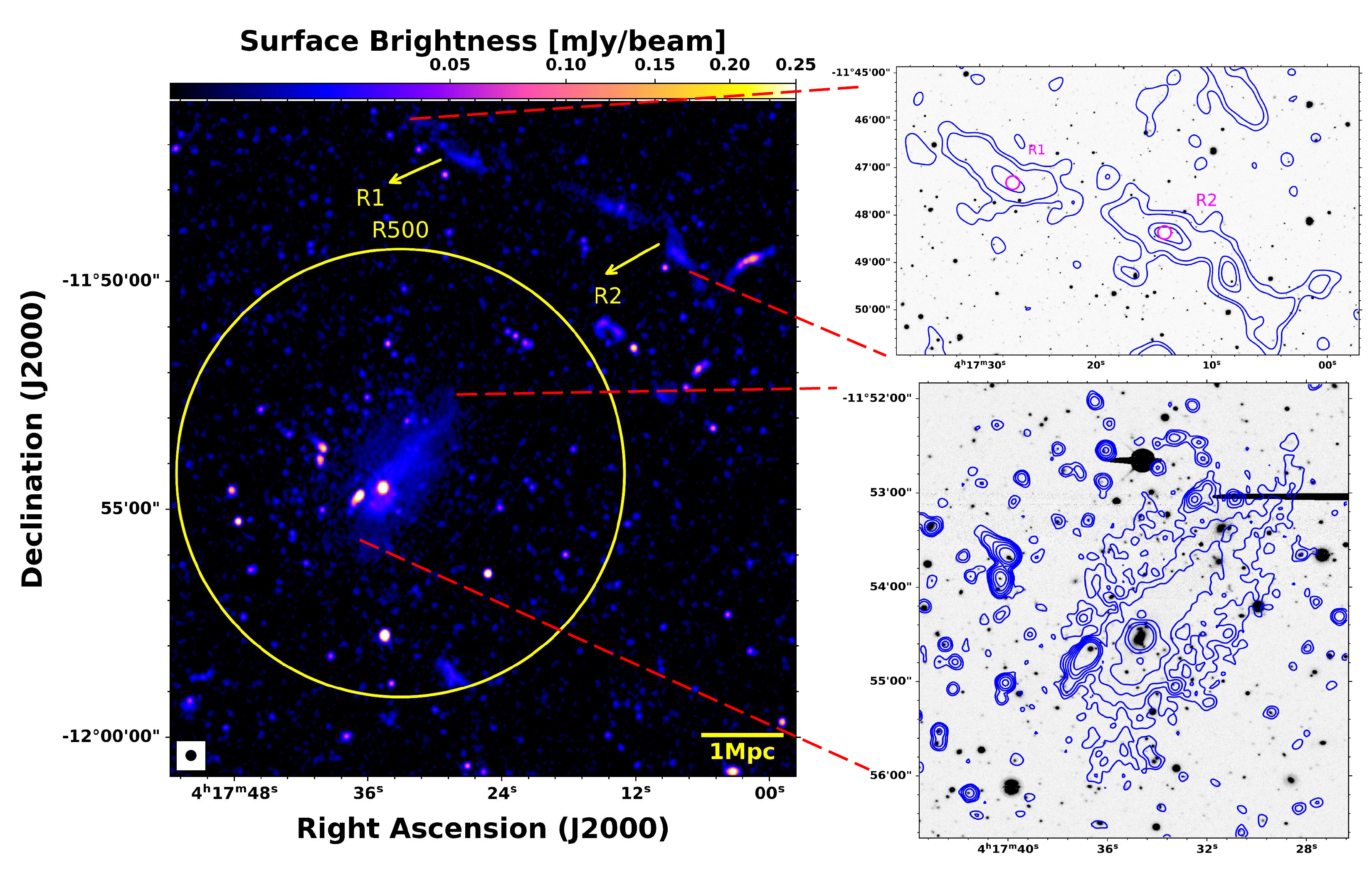}
    \caption{\textit{Left:} The MeerKAT full resolution 15$'$ $\times$ 15$'$ field of the MACSJ0417 is shown. The resolution of the image is 8$''$ $\times$7$''$. The radio halo is seen at the central region, and the yellow circle indicates the r$_{500}$ of the cluster, and at the peripheries, extended emission, R1, R2 (\revone{yellow arrows}) are labelled. \textit{Upper right:} DESI i-band image (greyscale) cutout of the MACSJ0417 field, focusing near the R1 and R2 region, is shown, overlaid with the radio emission at 1280 MHz, with a beam size of 15$''$ $\times$ 15$''$(IMG9). The magenta circle indicates the position of the peak of the radio emission. \textit{lower right:} The DESI i-band image of the cluster central part is shown, with overlaid contours from the MeerKAT full-resolution image (IMG7).}
    \label{macsj0417_intro_img}
\end{figure*}

\subsection{XMM-Newton}\label{xmm_data_analysis}

We used the deepest \textit{XMM-Newton} pointing of MACSJ0417 (obs-id: 0827011501, 0827310101) performed with the European Photon Imaging Camera (EPIC). A list of effective exposure times associated with each EPIC focal instrument is reported in Table~\ref{xmm_exposures_tab}. Our data reduction broadly followed the procedure described by \citet{eckert25}. First, we created calibrated event files using the standard screening process. We then applied the \texttt{mos-filter} and \texttt{pn-filter} tools to generate light curves for the \texttt{MOS1}, \texttt{MOS2}, and pn detectors of the EPIC, excluding intervals contaminated by soft-proton flares. The residual soft-proton contribution was estimated using the empirical relation between in- and out-of-field high-energy count rates and the soft-proton normalisation \citep{salvetti17}. For each instrument, we derived photon-count images, instrumental background estimates, and exposure maps across several energy bands. The particle background was modelled with \texttt{mos-spectra} and \texttt{pn-spectra}, scaling the filter-wheel-closed (FWC) data from the calibration database to match the observed high-energy background. This scaling was obtained by comparing the count rate in the unexposed detector corners with that of the FWC observations. Exposure maps were generated using \texttt{eexpmap} to correct for vignetting, chip gaps, and dead pixels, thereby providing the effective exposure at each detector position. Photon images, background maps, and exposure maps from all EPIC detectors and ObsIDs were then combined to produce the full \textit{XMM-Newton} image.

\begin{figure*}
    \centering
    \begin{tabular}{ccc}
        \includegraphics[width=0.38\textwidth]{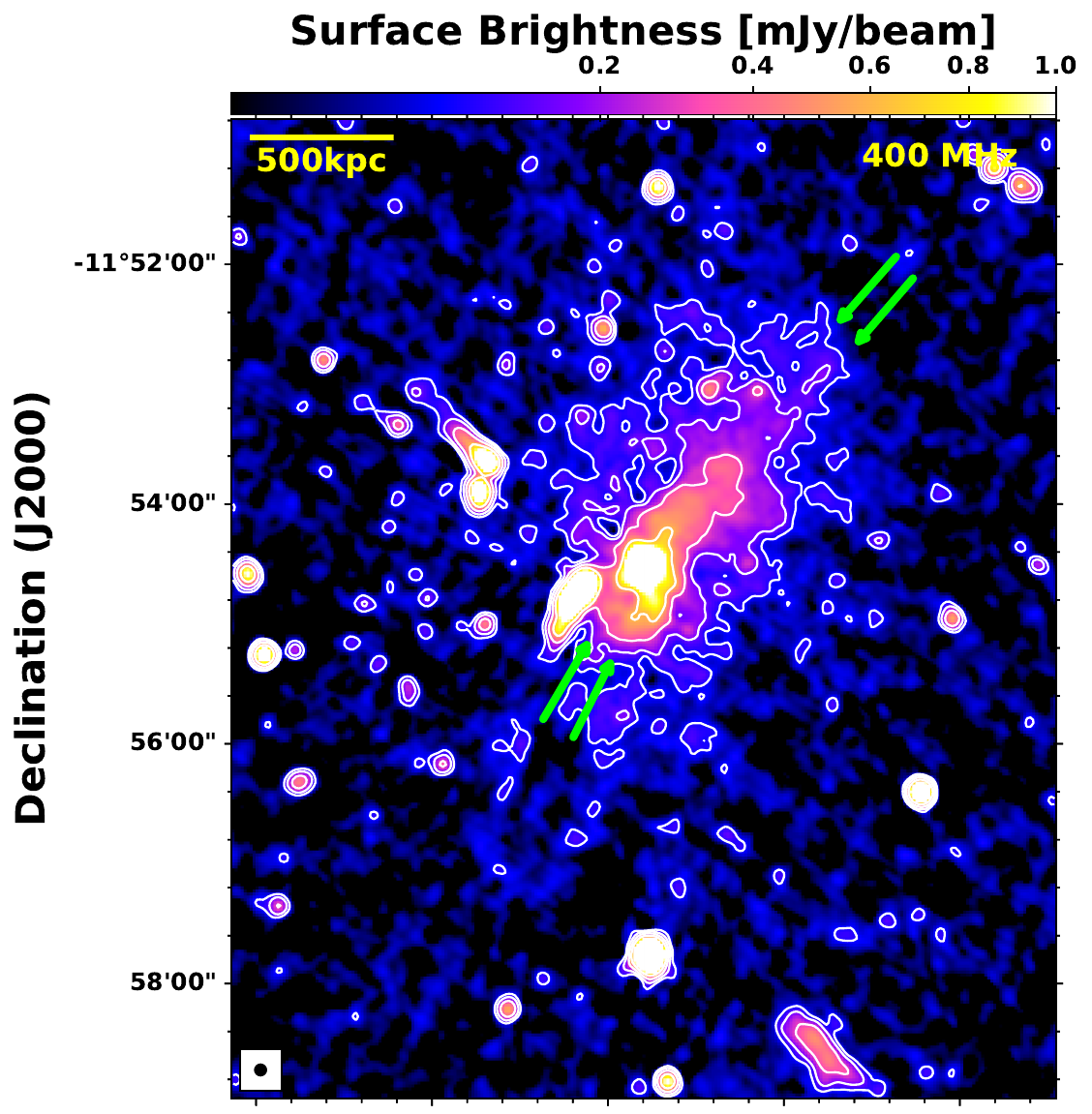} &
        \includegraphics[width=0.31\textwidth]{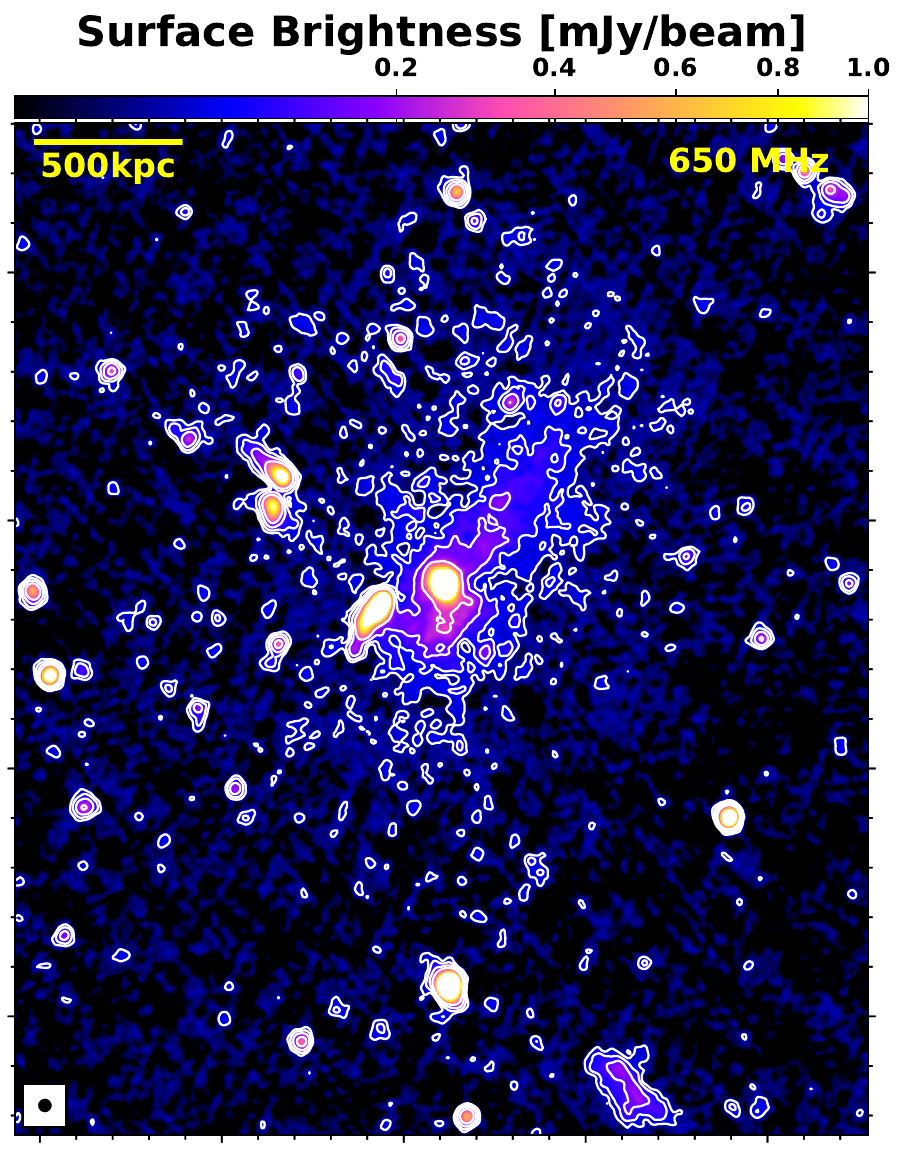} &
        
         \includegraphics[width=0.31\textwidth]{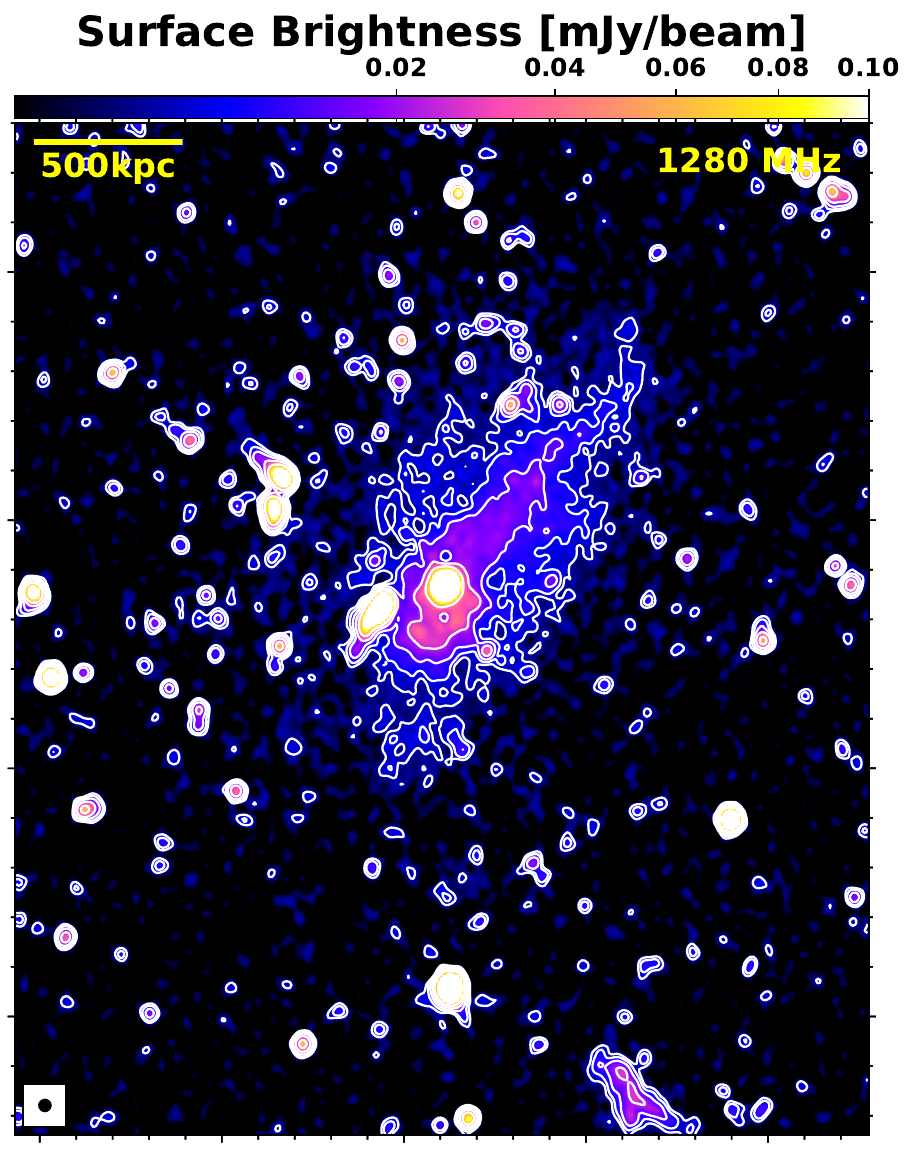}\\

        \includegraphics[width=0.38\textwidth]{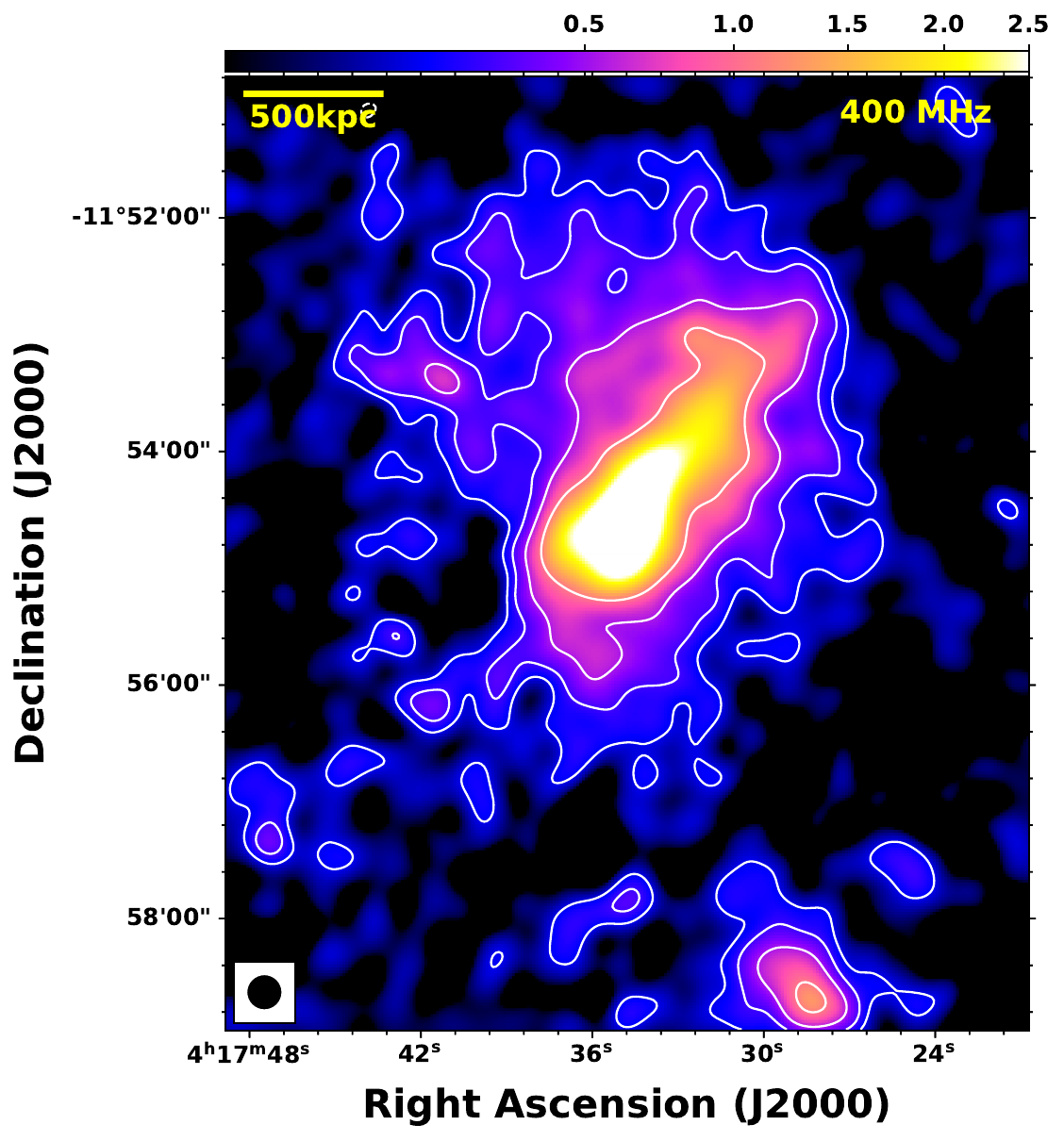} &
        \includegraphics[width=0.31\textwidth]{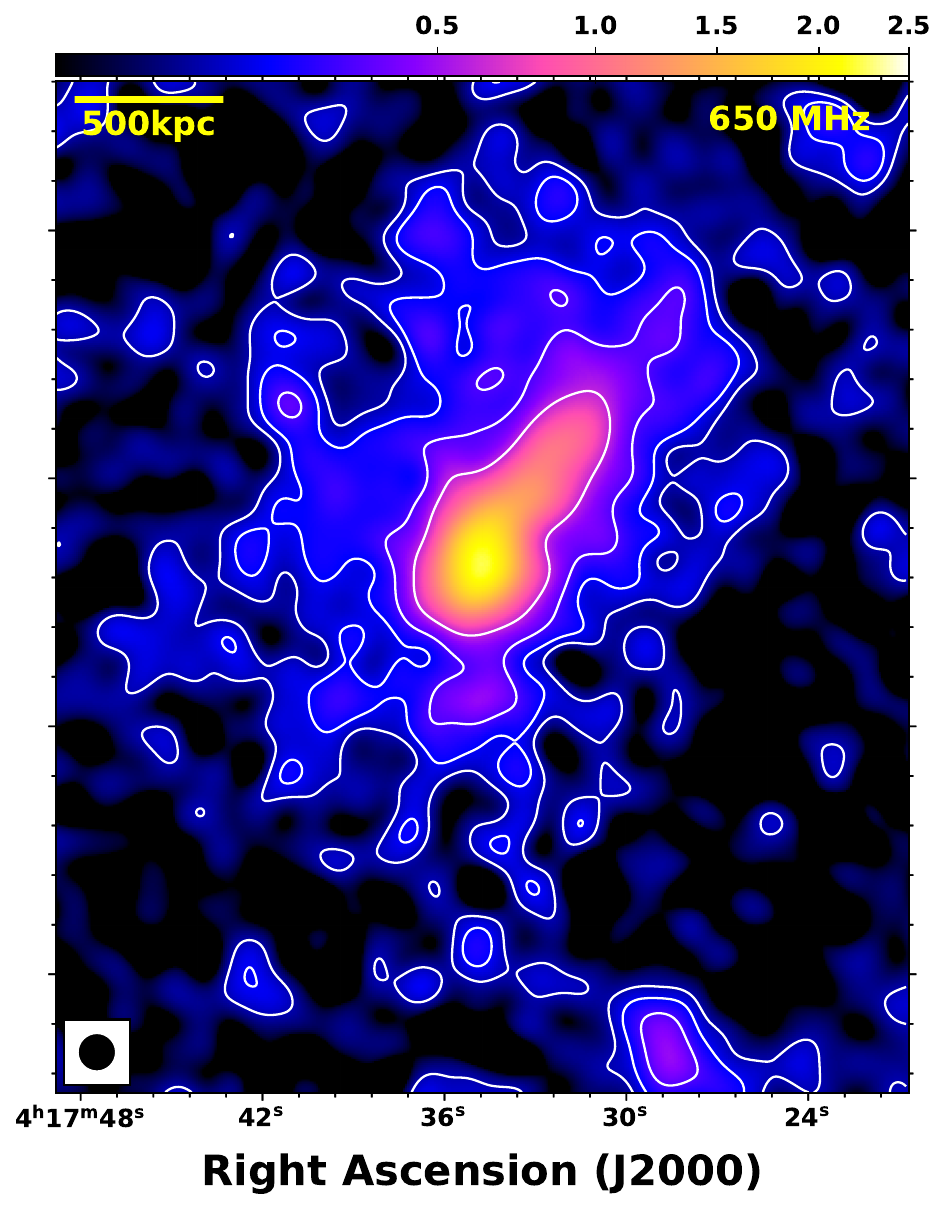} &
         \includegraphics[width=0.31\textwidth]{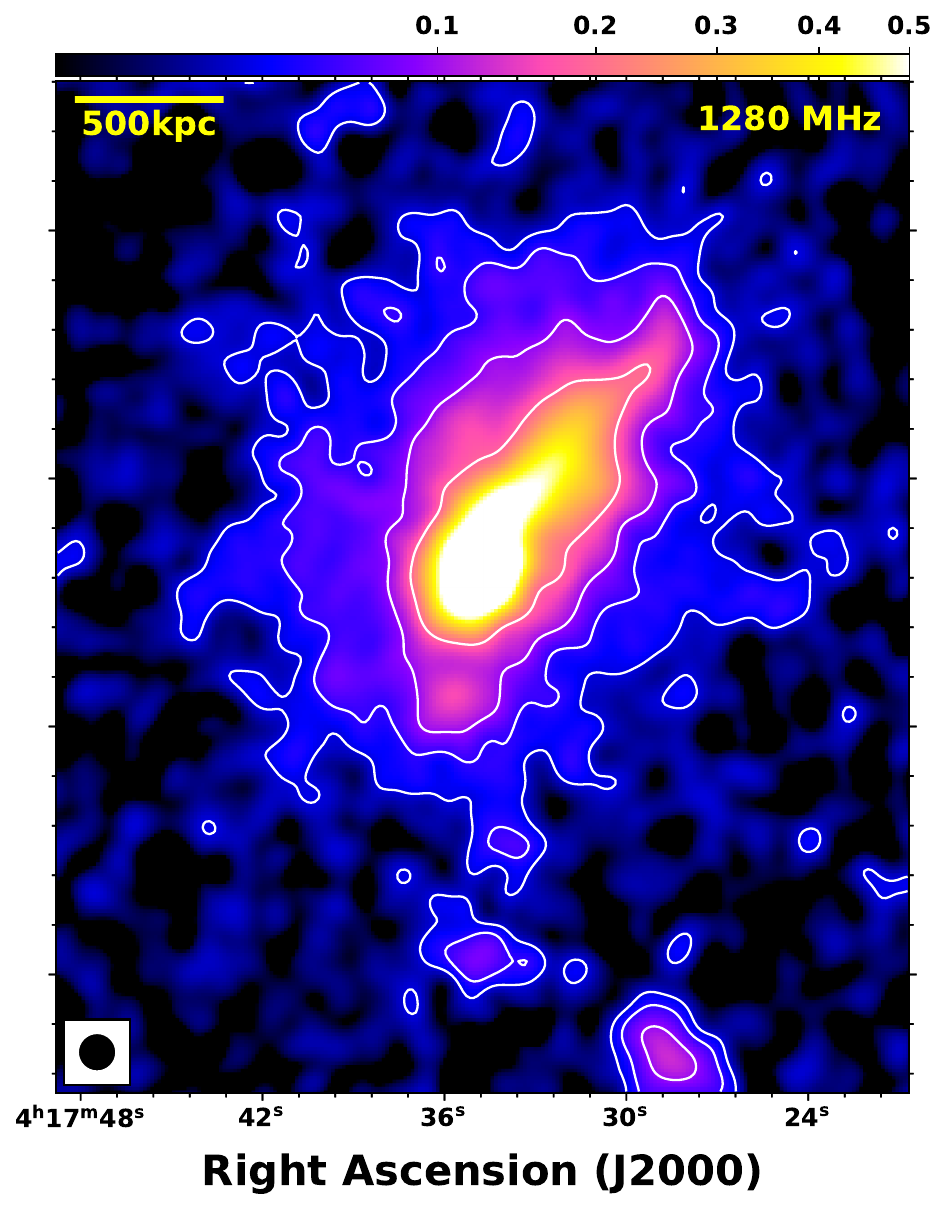}\\
    \end{tabular}
    \caption{High (upper) and low (bottom) resolution uGMRT band 3 (left), uGMRT band 4 (middle), and MeerKAT (right) images of the radio halo are shown. The green arrows in the upper left panel indicate the surface brightness edges. The low-resolution images are created at a common 15$''$ resolution, after subtracting the discrete sources. The contour levels start with $3\sigma_{\rm rms} \times [1,2,4,\dots]$, where the rms value for the images is reported in Table~\ref{img_summary}.}
    \label{macsj0417_gmrt_mkt}
\end{figure*}

\section{Radio Analysis} \label{radio_analysis}

\subsection{The diffuse emission}\label{radio_images}

We have obtained the most sensitive image of the MACSJ0417 cluster at frequencies below 1 GHz using the wide bandwidth capabilities of the uGMRT. We achieved noise levels of $\sigma_{\rm rms} = 27\mu$Jy~beam$^{-1}$ at 400 MHz and $9.0\mu$Jy~beam$^{-1}$ at 650 MHz. Figure~\ref{macsj0417_intro_img} displays the diverse nature of diffuse radio emission in the MACSJ0417 field. A detailed analysis of their observed properties and comparison with the X-ray morphology of the ICM is provided in the following sections.

\subsubsection{Radio halo}

The multi-frequency images of the extended emission are shown in Figure~\ref{macsj0417_gmrt_mkt}. Previous low-frequency observations at 235 MHz, limited by sparse \textit{uv}-coverage at short baselines, were unable to map the full extent of the radio halo \citep{2017MNRAS.464.2752P}. In contrast, our deep observations capture the emission out to scales exceeding 1 Mpc. The high sensitivity and angular resolution of these data allow us to characterise the halo in unprecedented detail. Figure~\ref{macsj0417_intro_img} shows the central radio halo together with two peripheral diffuse components, R1 and R2. In addition, a few discrete sources are detected; one of them, likely a tailed radio galaxy, exhibits a slight extension to the south and lies embedded within the diffuse halo, contributing $\sim$15\% of its total flux density. \revone{The radio halo is elongated along the SE--NW axis, while the low-frequency uGMRT and MeerKAT images also reveal extensions toward the east and west.} No filamentary substructures are evident, unlike those reported in some other halos (e.g., A2744, \citealt{rajpurohit21a}; A523, \citealt{vacca22}). As evident from Figure~\ref{macsj0417_gmrt_mkt}, the halo appears more extended at low frequencies. The largest linear sizes measured are 1.75 Mpc at 400 MHz, 1.48 Mpc at 650 MHz, and 1.4 Mpc at 1280 MHz. While halos generally appear smaller at higher frequencies due to spectral steepening in the outskirts, the sensitivity of MeerKAT allows detection of emission on larger scales, offering a more complete view of the morphology. The bright central region coincides with the merger axis, with surface brightness gradually declining outward. Notably, sharp edges (green arrows in Figure~\ref{macsj0417_gmrt_mkt}) are visible at the NW and SE outskirts, and their characteristics will be discussed in the following sections.

\subsubsection{Candidate relics}

In addition to the central diffuse halo, we detect elongated diffuse sources at the northern outskirts with peculiar morphologies. R1 lies to the north, while R2 extends along the NW–SE axis (Figure~\ref{macsj0417_intro_img}), with an arc-like shape. The emission is not clearly linked to any discrete source, though a compact source coincides with R2. No optical counterparts are seen in the Dark Energy Survey Instrument (DESI) i-band image (Figure~\ref{macsj0417_intro_img}). After subtracting compact sources and convolving to 15$''$, the projected sizes are $\sim$2.5$'$ (780 kpc) for R1 and 4.55$'$ (1400 kpc) for R2. Both sources are situated at $\sim$2.9 Mpc (1.7r$_{500}$) from the cluster centre. \citet{knowles22, kolokythas25} proposed R1 and R2 as candidate relics, but the absence of detected shocks and their large distance ($\sim r_{\rm vir}$) make this uncertain.

\subsection{Integrated spectrum}\label{sec:int_spec}

\begin{figure}
\includegraphics[width=\columnwidth]{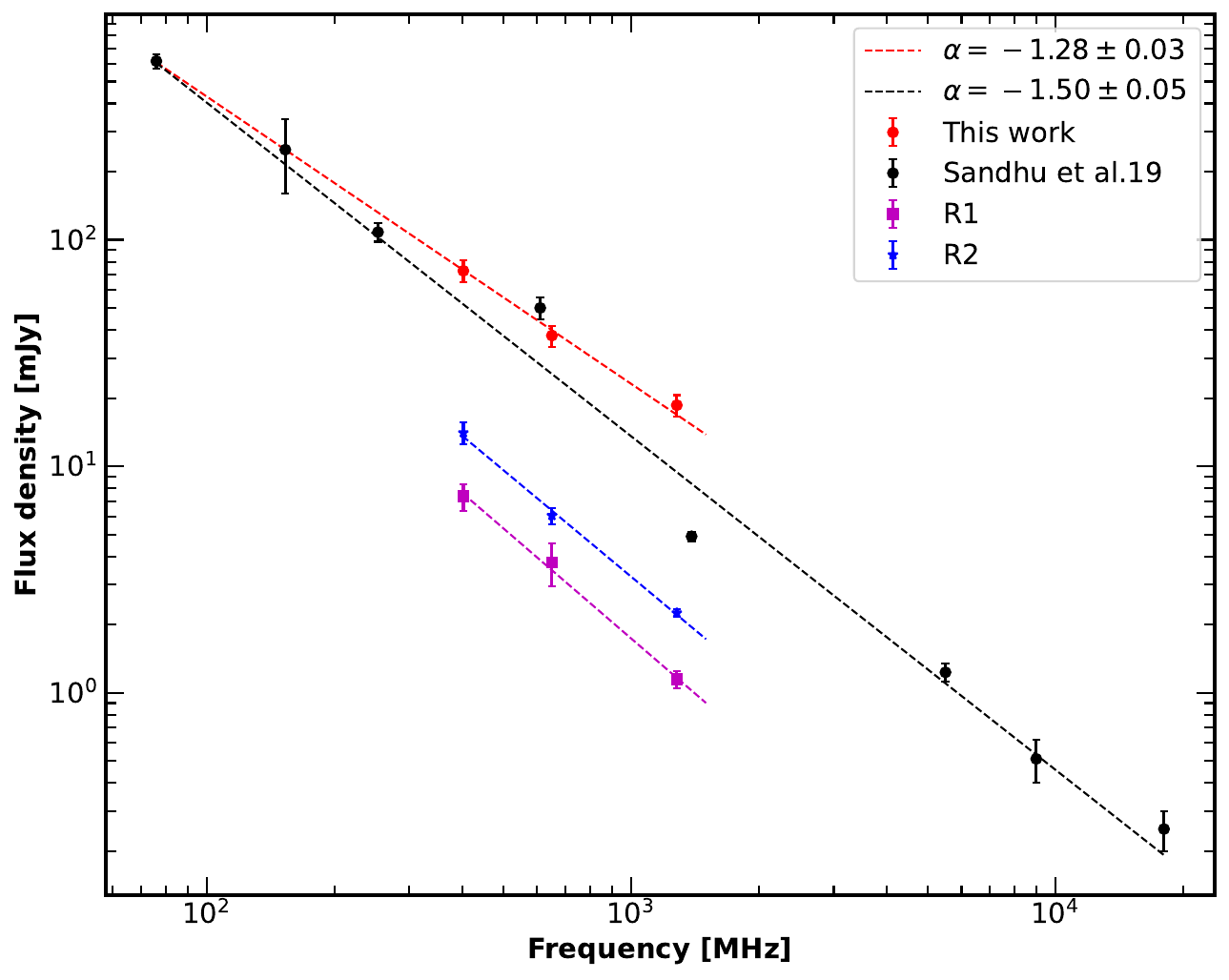}
    \caption{The integrated spectrum of the radio halo and candidate relics is shown. The red dashed line shows the spectrum obtained using our work, and the black dashed line shows the spectrum from \citet{2019JApA...40...17S}. The blue and the magenta lines show the integrated spectrum of R2 and R1, respectively.} 
    \label{halo_int_spec_img}
\end{figure}

The spectral shapes of the radio halos remain poorly constrained due to challenges in measuring extended emission, due to missing short \textit{uv}-baselines, deconvolution issues, and contamination from unrelated sources. We measured the flux densities of the halo and candidate relics and derived their integrated spectral index by combining our results with literature values. \revone{The integrated flux densities were measured from the compact-source-subtracted images. To ensure a consistent comparison across frequencies, we used the IMG3, IMG6, and IMG9 images at all frequencies, which were produced with a common restoring beam of $30''\times30''$, and an inner \textit{uv}-cut of 0.2 k$\lambda$. The flux densities were measured within identical source regions at all frequencies}. The radio halo spectrum is well described by a single power law between 400--1280 MHz, with an integrated index of $-$1.28$\pm$0.03 (Figure~\ref{halo_int_spec_img}), consistent with measurements below 200 MHz. \citet{2019JApA...40...17S} reported a steeper index of $-$1.5 between 76 MHz$-$18 GHz. Our result likely reflects observational differences: their fit relies on high-frequency ATCA data (5.5$-$18 GHz) with limited short \textit{uv}-coverage, which can underestimate diffuse emission and steepen the spectrum artificially. In addition, our high-resolution 1280 MHz imaging provides more accurate compact-source subtraction, while their spectrum combines heterogeneous datasets with differing resolutions, sensitivities, and \textit{uv-}coverages. These factors plausibly explain the discrepancy and underline how \textit{uv}-coverage and frequency sampling affect spectral index estimates of extended sources.

\begin{table}
  \centering
  \caption{Flux density estimates for the extended sources.}
\begin{tabular}{@{}cccc@{}}
    \hline
     Source & Freq. & Flux density (mJy) & Ref. \\
      \hline\hline
     &76 & 615.0 $\pm$ 44.0 & \citet{2019JApA...40...17S}   \\
    
    &150 & 250.0 $\pm$ 90.0 & \citet{2019JApA...40...17S}   \\
    
    Halo  & 235 & 108.1 $\pm$ 10.0 & \citet{2019JApA...40...17S} \\
    
    &402 & 73.0 $\pm$ 8.0 & This work \\
    
    &650 & 36.4 $\pm$ 4.0 &  This work\\
    
    &1280 & 18.6 $\pm$ 2.0 & This work \\

    \hline 
    
      & 400 & 7.4 $\pm$ 0.8& This work  \\
    R1 & 650 & 3.8 $\pm$ 0.4& This work \\
    & 1280 & 1.1$\pm$ 0.1& This work \\
    \hline

      & 400 & 14.1 $\pm$ 1.4& This work \\
    R2 & 650 & 6.1 $\pm$ 0.6& This work \\
    & 1280 & 2.3$\pm$ 0.3& This work \\
    \hline

\end{tabular}
    
  \label{int-spec-tab}
\end{table}

Our measurements for R1 and R2 result in a spectral index of $-$1.60$\pm$0.05, indicating that they exhibit similarly steep spectra. Assuming the \revone{Diffusive Shock Acceleration} (DSA) \citep{Blandford1978, Bell1978, Bell1978a} to be the origin of the particle acceleration in the relic region, the Mach number (M$_{\rm S}$) is related to the integrated spectral index $\alpha_{\rm int}$ as 

\begin{equation}\label{eq.4}
    \rm M_{\rm S} =\sqrt{\frac{\alpha_{\rm int} +1}{\alpha_{\rm int} -1}} = 2.08 \pm 0.08
\end{equation}

The obtained Mach number is consistent with a scenario of weak merger shocks propagating outwards. The spectral indices of the `accretion relic' have been predicted to have comparatively flatter spectral indices ($\sim-$ 1), than the standard relics \citet{hoeft07}. The $k$-corrected (extrapolated) radio power of R1 (P$_{\rm 1.4 GHz} = (1.04 \pm 0.12) \times 10^{24}$W~Hz$^{-1}$) and R2 (P$_{\rm 1.4 GHz} = (5.98 \pm 0.6) \times 10^{23}$W~Hz$^{-1}$) fits well in the radio power vs. linear size relation (with high scatter) of other known relics \citep[e.g.,][]{balboni25}.

\subsection{Radio surface brightness profile} \label{SB_prof_halo}

The surface brightness profile (assuming spherical symmetry) of the radio halos is generally explained as :

\begin{equation} \label{halo_model_eq}
    \rm I(r) = \rm I_{0} e^{-\frac{r}{r_{e}}}
\end{equation}

where I$_{0}$ is the central surface brightness and r$_{e}$ is the e-folding radius, at which the surface brightness is I$_{0}$/e. Although the exponential model provides a good description of radio halo profiles \citep[e.g.,][]{murgia09}, observations with new-generation facilities reveal substructures and multiple components, highlighting its limitations in capturing the complex and irregular morphology of radio halos \citep[e.g.,][and references therein]{2021A&C....3500464B}. We estimated the average surface brightness in concentric elliptical annuli centred on the radio peak, with widths of half the image beam size (15$''$). \revone{As a result, adjacent radial bins are not fully independent, although this is not expected to significantly affect the derived exponential-profile parameters.} Residuals of discrete sources were masked, and only regions above $3\sigma_{\rm rms}$ were considered. The uncertainties in the surface brightness account for both the flux density scale error, statistical noise, and discrete source subtraction error.

\begin{figure}
\includegraphics[width=\columnwidth]{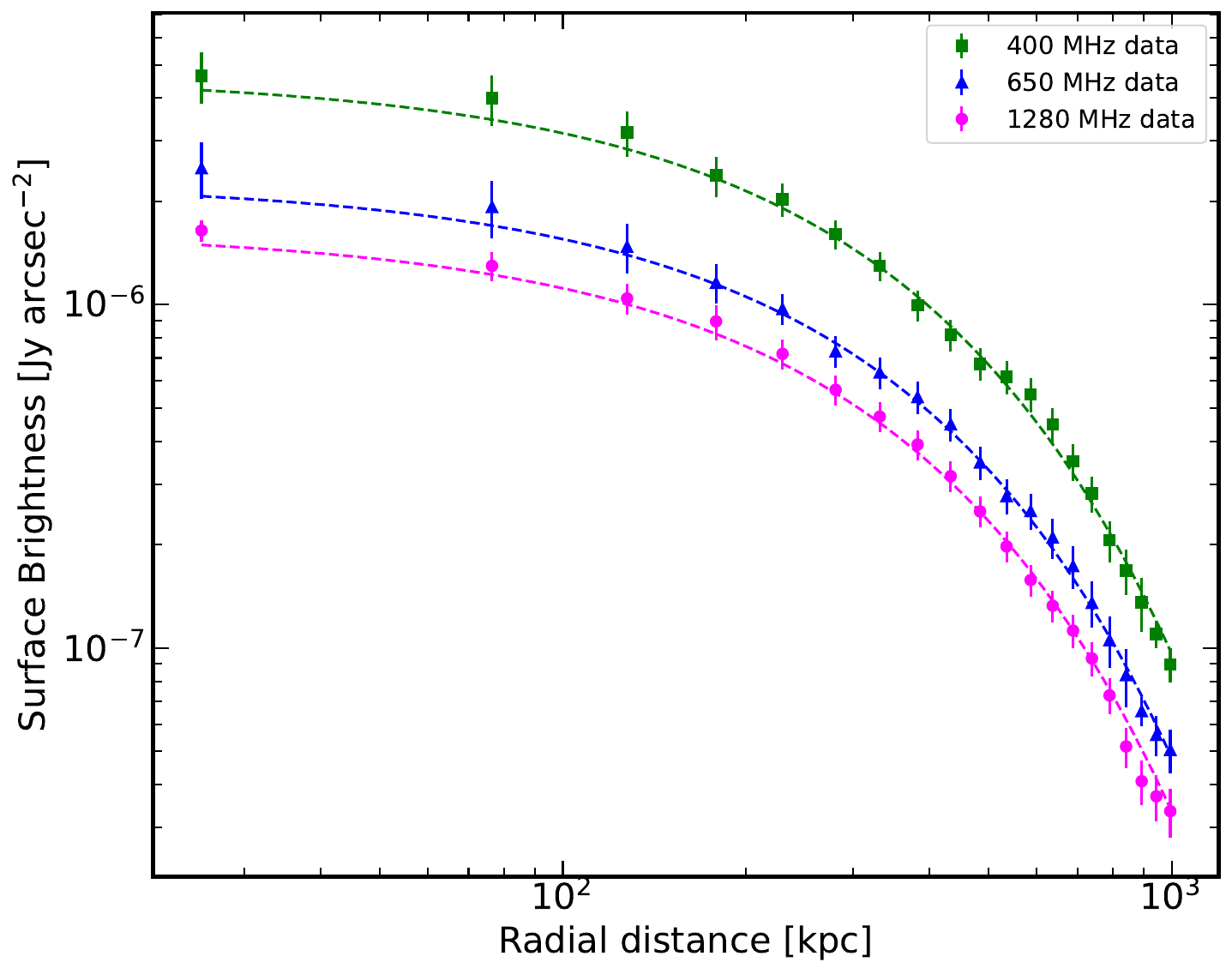}
    \caption{The radial surface brightness profile for the halo
    emission is shown for the three different frequencies of our study. The dotted line shows the model (equation~\ref{halo_model_eq}) fitted to the radial profile for the total halo emission. Each data point indicates the average radio surface brightness from each annulus.} 
    \label{halo_Sb_profile_img}
\end{figure}

In Figure~\ref{halo_Sb_profile_img}, we present the radial surface brightness profiles of the radio halo at each frequency. The best-fit parameters of the equation~\ref{halo_model_eq} are reported in Table~\ref{radio_SB_fit_param}. The radio halo emission is well described by a single exponential profile, with no secondary components or distinctive features in the surface
brightness profile. The best-fit e-folding radius varies over the frequency, larger at lower frequency and smaller at higher frequency, indicating a more peaked profile of the radio emission at higher frequencies. We have also estimated the emissivity of the halo to be $\sim$10$^{-42}$ erg~s$^{-1}$, at 1.4 GHz using equation 2 of \citep{rajpurohit25}. The estimated central I$_{0}$ and e-folding radius is well fitted in the I$_{0}$ $-$ r$_{e}$ plane of the giant radio halos \citep[Sample from][]{murgia09,rajpurohit25}, irrespective of the fact that it is situated at a moderately high redshift. \citet{murgia09} reported a tight correlation between the total radio power of halos and their e-folding radius, P$_{1.4}$ $\propto$ r$_{e}^{3}$, and MACSJ0417 shows good agreement with the properties of previously known classical halos.  

\begin{table}
  \centering
  \caption{Fitting results obtained from an exponential fit}
  \begin{tabular}{@{}cccc@{}}
    \hline
      Freq & $\chi_{r}^{2}$ & I$_{0}$ & r$_{\rm e}$  \\

      (MHz) & & ($\mu$Jy~arcsec$^{-2}$) & kpc. \\
    \hline\hline
    
    400 &  1.21 & 4.66 $\pm$ 1.65 & 300.8 $\pm$ 10.0  \\

    650 &  1.15 & 2.28 $\pm$ 0.09 & 268.0 $\pm$ 11.3  \\

    1280 &  1.19 &1.64$\pm$ 0.04 & 230.0$\pm$ 9.2  \\

    \hline
    \end{tabular}
    \label{radio_SB_fit_param}
\end{table}

\subsection{Detection of radio surface brightness edge} \label{SB_edge_detection}

\begin{figure}
\includegraphics[width=\columnwidth]{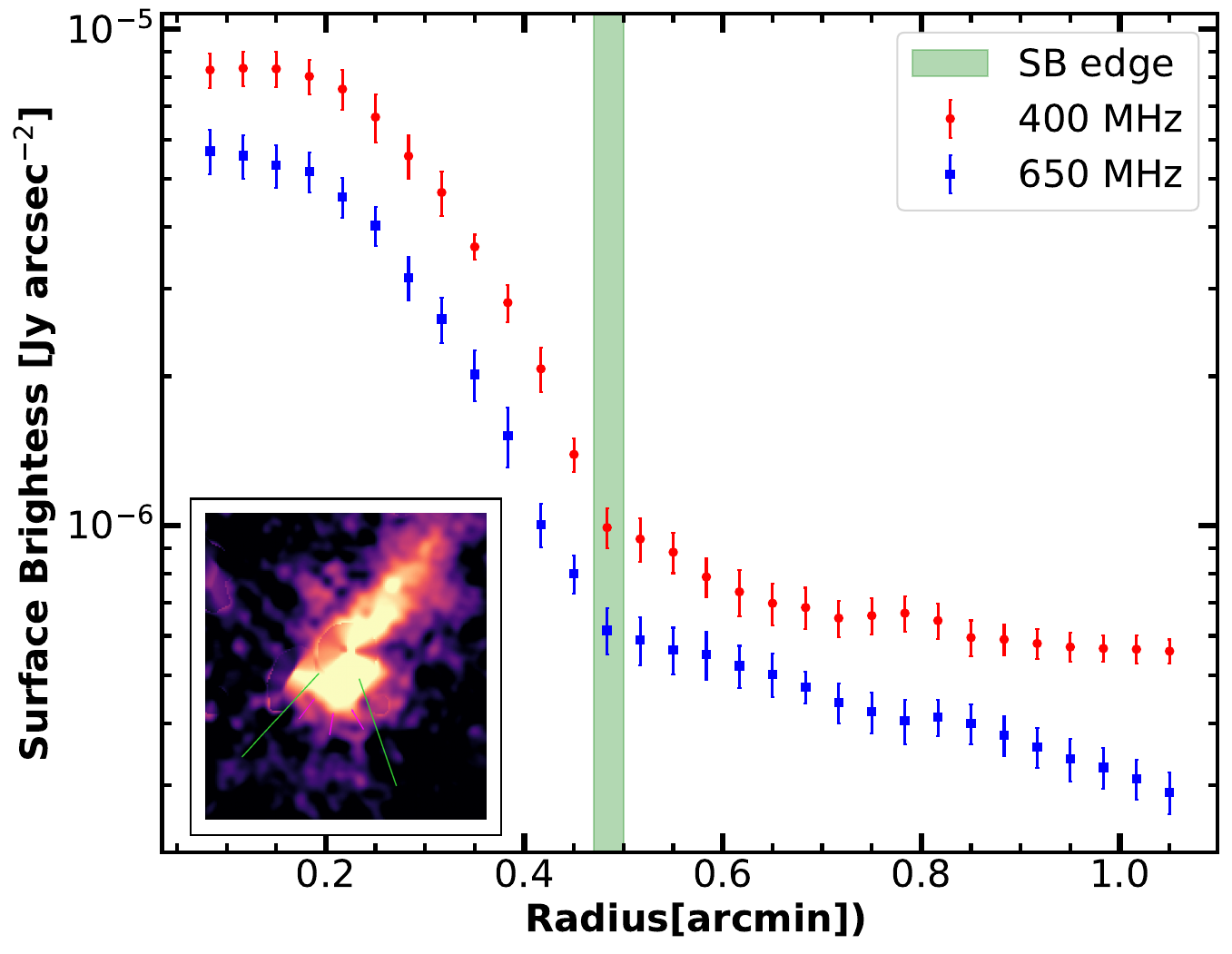}
    \caption{Radio surface brightness profile extracted from the point source-subtracted image is shown. The green coloured segment in the inset panel denotes the sector used to extract the surface brightness profiles plotted following the different colours for different frequencies. The green shaded region indicates the location of the edge detected in \citet{botteon23}.} 
    \label{SB_edge_halo}
\end{figure}

Dynamical motions in the ICM can leave sharp features, which, with sufficient resolution and sensitivity, can also be detected in radio surface brightness maps. \citet{botteon23} reported such a radio discontinuity in MGCLS\footnote{\url{http://mgcls.sarao.ac.za/}}
 clusters, coincident with an X-ray edge, highlighting the link between thermal and non-thermal components. Our deep uGMRT images enable a search for possible discontinuities in the diffuse halo emission. \citet{botteon23} reported a sharp surface brightness edge in the southeastern region, with the brightness dropping by a factor of 5 within $\sim$100 kpc. Using our uGMRT images at comparable resolution (9$''\times$9$''$), we searched for the same feature following their procedure (1$''$ radial binning). Figure~\ref{SB_edge_halo} shows the surface brightness profiles from the discrete source-subtracted image, revealing a decline by a factor of $\sim$6–7 within $\sim$130 kpc, consistent with the MeerKAT result. The discontinuity appears at a similar location in all three frequencies (400, 650, and 1280 MHz). A GGM-filtered image (Appendix~\ref{ggm_image}) further highlights this edge. A cold front was previously reported here \citep{pandge19,Botteon2018}, suggesting a strong link between thermal and non-thermal structures, though the detailed modelling and spectral constraint across the edge is beyond the scope of this work.

\subsection{Resolved spectral map} \label{res-spec-map}

\begin{figure*}
    \includegraphics[height = 16cm, width = 0.93 \textwidth]{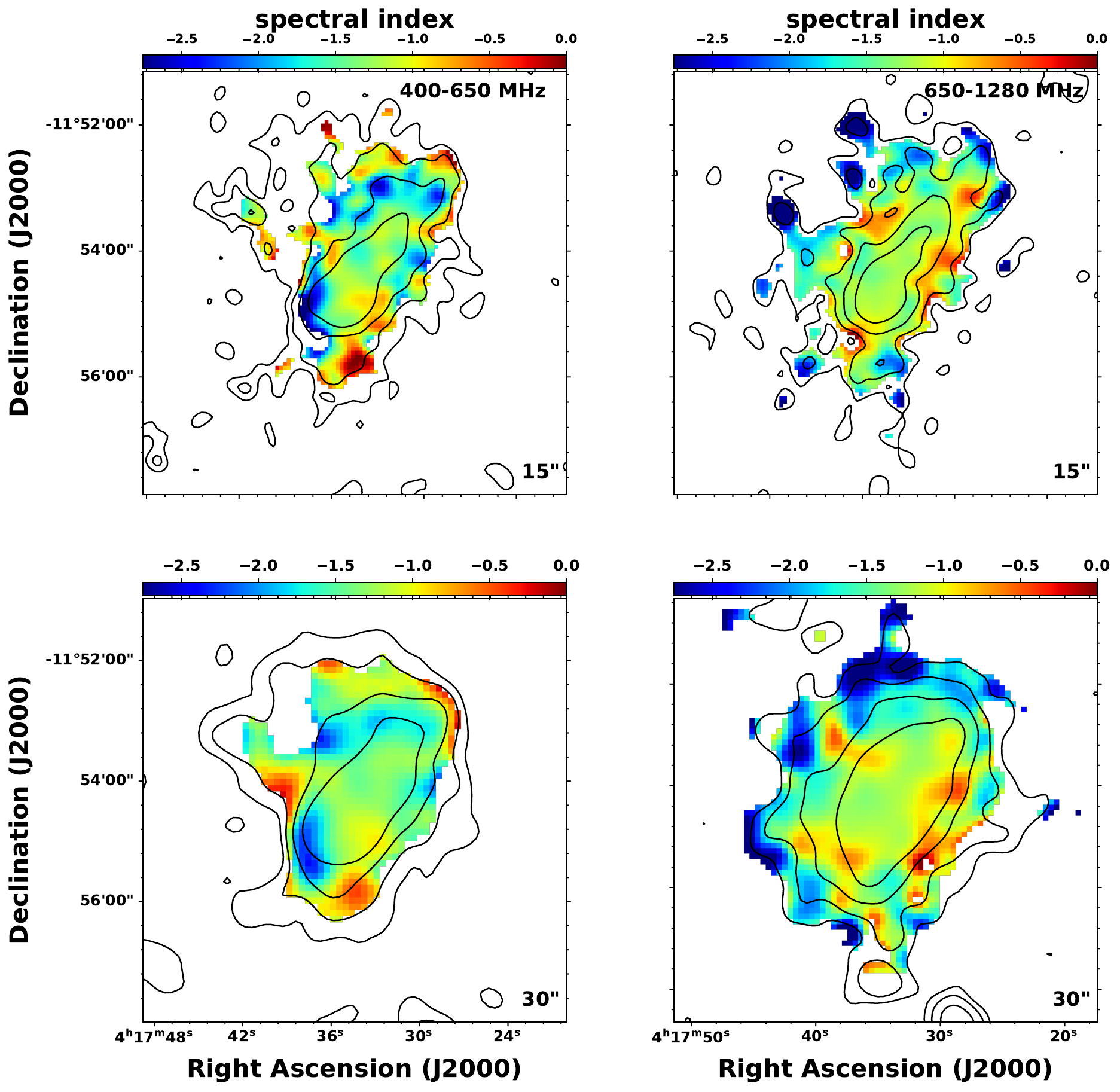}
     
    \caption{\textit{Upper left:} The spectral index map between the 400 and 650 MHz is shown at a 15$''$ resolution. The black contours are from the uGMRT 400 MHz image at 15$''$ resolution at 3$\sigma_{\rm rms}$ $\times$ [1,2,4], with $\sigma_{rms} = 45 \mu$Jy~beam$^{-1}$. \textit{Upper right:} The spectral map between 650 and 1280 MHz is shown at 15$''$ resolution. The black contours are from the uGMRT 650 MHz image at 15$''$ resolution at 3$\sigma_{\rm rms}$ $\times$ [1,2,4], with $\sigma_{rms} = 28.7 \mu$Jy~beam$^{-1}$. \textit{Lower left:} Resolved spectral map for the uGMRT frequencies is shown at 30$''$ resolution, with black contours overlaid from the uGMRT 30$''$ resolution image at 400 MHz. \textit{Lower right:} The spectral map at 30$''$ resolution between 650 and 1280 MHz, with black contours overlaid from the MeerKAT 30$''$ resolution image.}
    \label{spec_maps}
\end{figure*}

\begin{figure}
\includegraphics[width=\columnwidth]{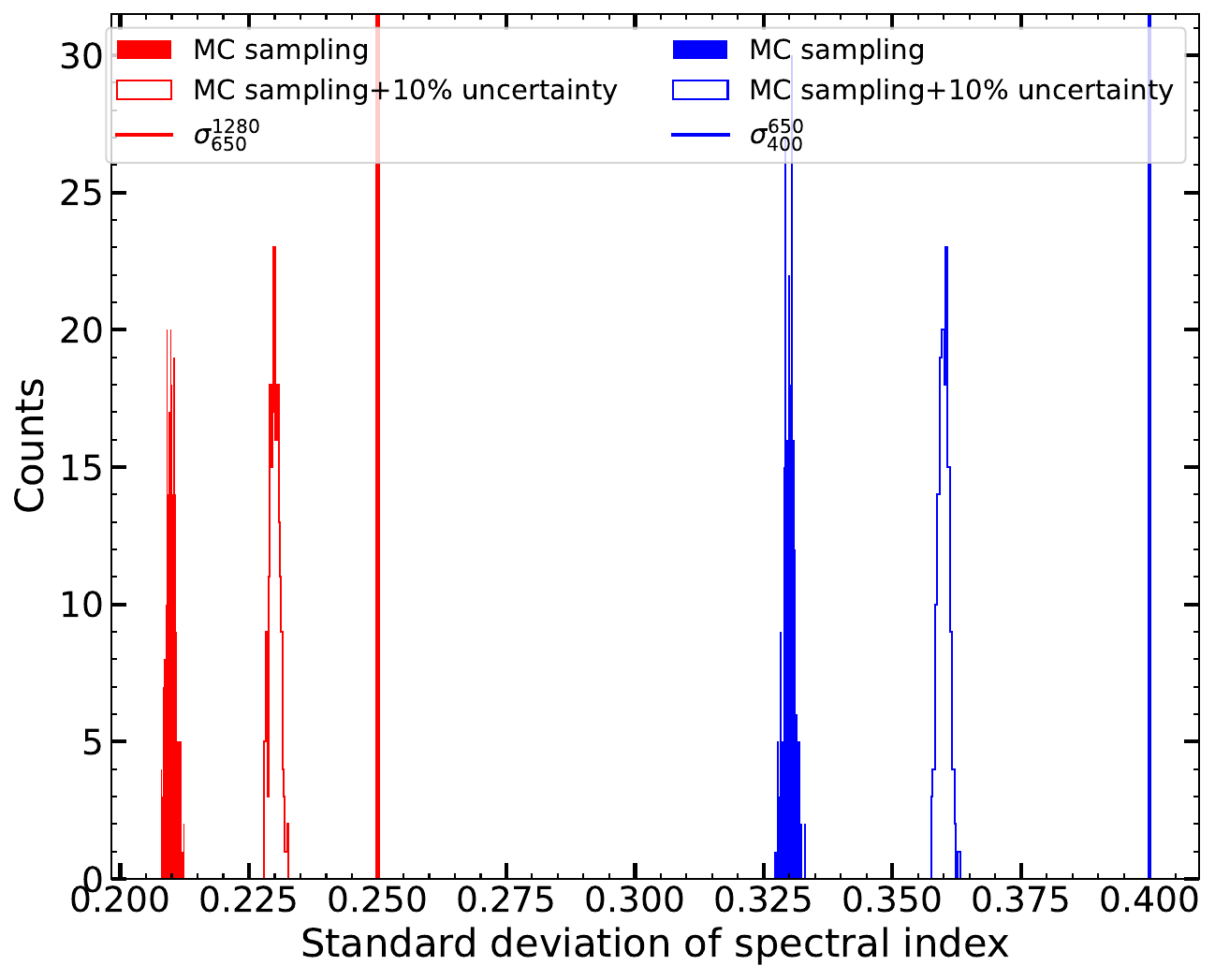}
    \caption{The observed standard deviation (vertical line) of the spectral index in the halo region, in comparison to the standard deviation expected purely from the thermal noise and with an additional uncertainty contribution from the calibration. The two different colours indicate the distribution from the two different frequency maps.} 
    \label{halo_statistics_img}
\end{figure}

\begin{figure}
\includegraphics[width=\columnwidth]{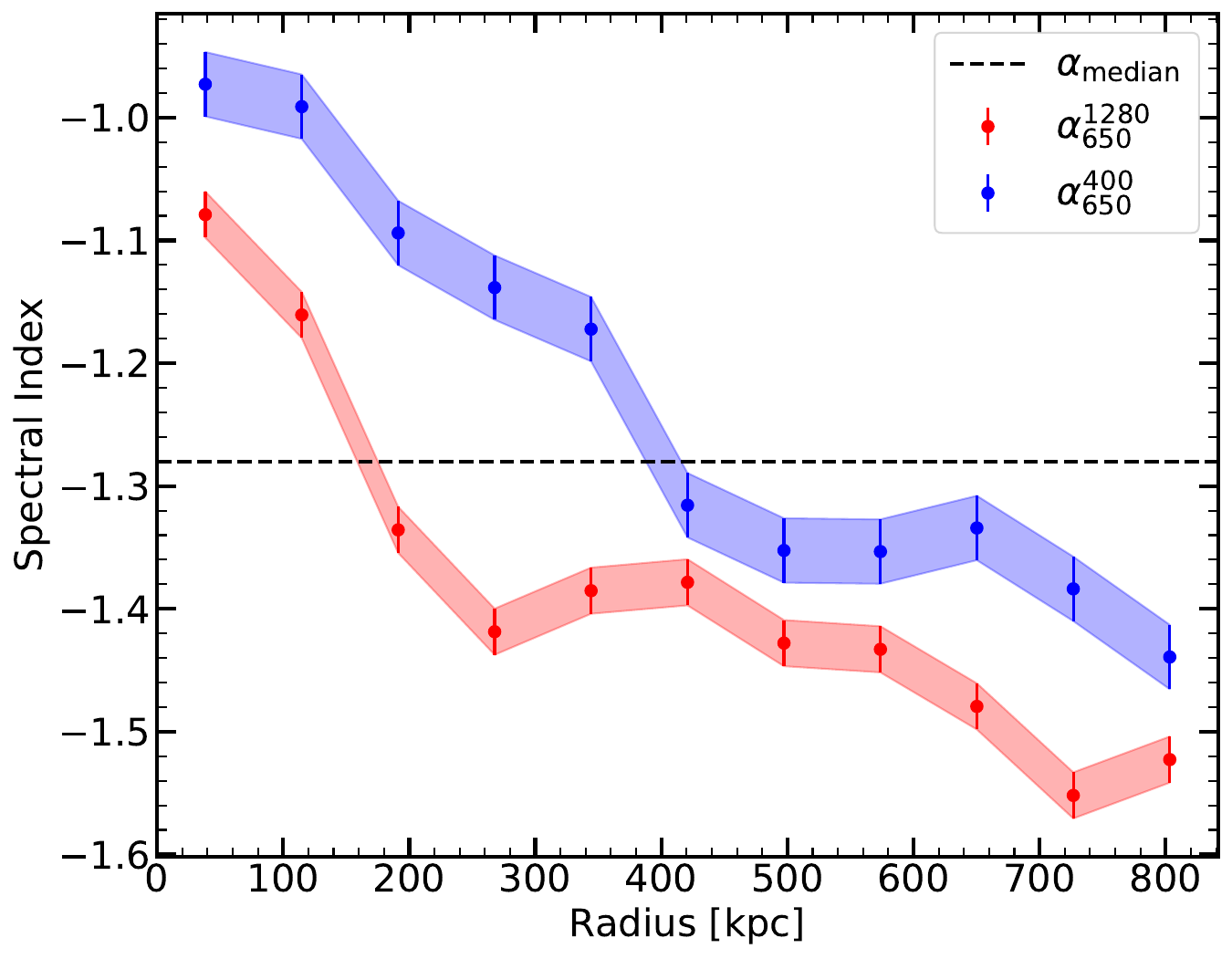}
    \caption{The radial profile for the average spectral index is shown for both the spectral index maps at 30$''$ resolution. The shaded region corresponds to a 1$\sigma$ error bar on measurements. The black dashed line shows the median spectral index distribution.} 
    \label{halo_Spec_profile_img}
\end{figure}

Using our uGMRT and MeerKAT observations, we made spectral index maps of the halo at angular resolutions of 15$''$ and 30$''$. Following imaging, we aligned the uGMRT and MeerKAT data on a common grid, and applied a 3$\sigma_{\rm rms}$ mask to both maps. Although this cutoff ensures the reliable measurements of the spectral indices, it may also exclude regions with intrinsically flat or steep spectra due to the different sensitivity of the telescopes. The astrometry accuracy between the two datasets revealed no significant offsets ($\sim$1$''$). The detailed procedure for estimating the spectral index and its corresponding error maps follows the method described in \cite{degasperin17}.

We have presented the spectral index maps at two different frequencies, 400--650 and 650--1400 MHz (Figure~\ref{spec_maps}), with the corresponding error map shown in Appendix~\ref{sec:spec_err_maps}. In the central region, flat spectral indices ($\sim -$0.70) are seen, which may be related to the residual of the central point source. The average spectral index is steep ($\sim -1.2$) along the NW-SE direction. The resolution ($15''$) of our maps prevents us from putting any constraint on the behaviour of the spectral index across the SE edge. \revone{Steep spectrum regions ($\alpha \lesssim -1.5$) are also observed in the cluster outskirts; however, the low surface brightness of the emission results in larger uncertainties, and these features are therefore detected at lower significance.} Due to the mixed spectral index distribution, we estimate the standard deviations in the halo region for both the spectral maps at 15$''$ resolution. The estimated standard deviation from the spectral maps is std($\alpha_{400}^{650}) = 0.41$ and std($\alpha_{650}^{1280}) = 0.25$, in comparison to the median uncertainties in the spectral index error map, which were calculated to be 0.35 and 0.21. To test whether this can be explained by measurement noise or instrumental effects, we performed Monte Carlo (MC) resampling of a uniform spectral index map using the pixel–wise uncertainties. \revone{The observed scatter exceeds simulation predictions (Figure~\ref{halo_statistics_img}). Even after adding a 10\% calibration-related error scaling with surface brightness, the simulations fall short. Matching the observed level requires a larger error fraction ($\sim$12--13\%). Potential residual errors associated with the subtraction of compact sources could also contribute to the observed spectral index fluctuations. Their impact is difficult to quantify precisely; however, regions with obvious subtraction residuals were excluded from the analysis, and the excess scatter is observed across a substantial fraction of the halo area rather than being confined to the vicinity of bright compact sources. While compact-source subtraction uncertainties may contribute locally to the spectral index distribution, they are unlikely to fully account for the observed level of fluctuations. The observed scatter therefore suggests statistically significant spectral index fluctuations and inhomogeneities within the synchrotron-emitting volume.}

We also examined the azimuthal variation of the average spectral index radially. The halo emission was sampled in concentric annuli centred on the cluster, and the resulting radial profile (Figure~\ref{halo_Spec_profile_img}) shows a clear steepening, from $\sim -1.0$ in the core to $\sim -1.5$ in the outskirts. Radial profiles from both spectral maps show this trend, although steepening occurs at different radii in the two cases. The frequency dependence of the e-folding radius further supports this behaviour, being larger at low and smaller at high frequencies. The steep spectral regions in the outskirts may correspond to areas of less efficient turbulence and/or weaker magnetic fields.

\begin{figure*}
    \includegraphics[width = 0.92 \textwidth]{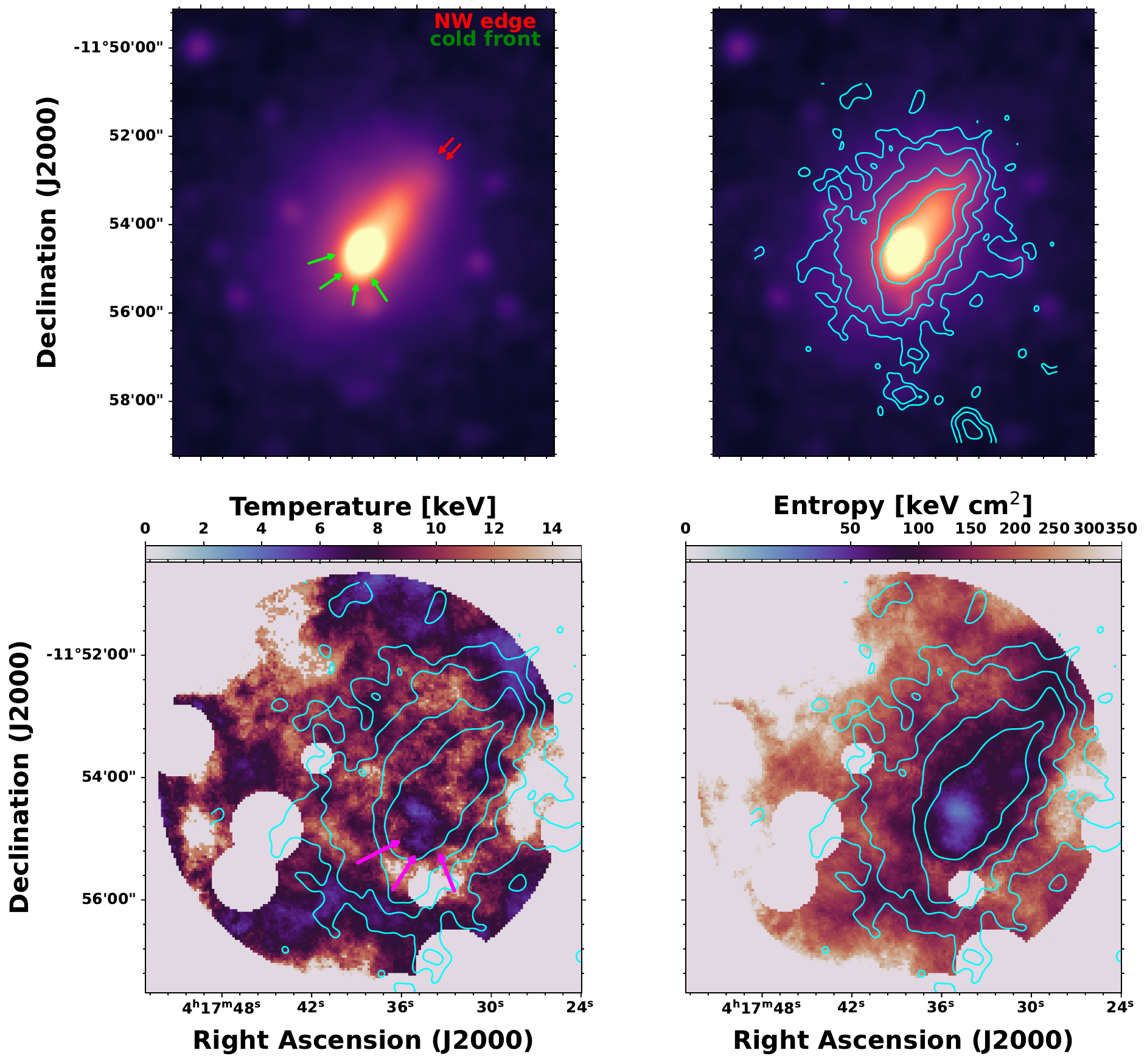}
     
    \caption{\textit{Upper left:} The exposure-corrected, background-subtracted \textit{XMM-Newton} is shown, smoothed with a Gaussian Full-width half maximum of 10$''$. The red and green arrows show the presence of the discontinuity in those locations. \textit{Upper right:} The same is shown, with an overlay of the radio emission from 1280 MHz. The contour level starts from 3$\sigma_{\rm rms}$ $\times$ [1,2,4...], with $\sigma_{rms} = 9\mu$Jy~beam$^{-1}$. \textit{Lower left:}temperature map extracted from the smoothed XMM-Newton surface brightness map is shown, overlaid with the radio emission at 1280 MHz (similar contour level to the previous). The magenta arrow shows the location of the SE cold front. \textit{Lower right:} Here, the entropy map for the same is shown, with overlay from the radio emission.}
    \label{xray_maps}
\end{figure*}

\section{X-ray analysis}\label{xray_analysis}

\subsection{Surface brightness map}

X-ray observations of MACSJ0417 reveal an asymmetric ICM morphology in the \textit{XMM-Newton} surface brightness map, with central diffuse emission extended along the SE–NW axis (Figure~\ref{xray_maps}, left). The diffuse emission is detected out to $\sim$1.9 Mpc, with comparable extents to the east and west, but more extended northward ($\sim$1.1 Mpc) than southward ($\sim$0.8 Mpc). Several discrete sources are embedded within the diffuse emission. A compact, bright core is evident, \delone{identical to} \revone{reminiscent of} a relaxed cool core cluster. The overall X-ray morphology appears comet-like, with the bright core suggesting a high-velocity encounter, similar to A2146 \citep[e.g.,][]{chadayammuri22}. \citet{pandge19} reported a tail-like feature toward the NE; however, no such structure is seen in our \textit{XMM-Newton} image. The \textit{Chandra} map revealed a surface brightness discontinuity, later confirmed as a cold front, which is also visible in the XMM data, possibly extending further northward. Another discontinuity in the NW, reported by \citet{Botteon2018}, is also confirmed in our \textit{XMM-Newton} map.

In Figure~\ref{xray_maps} (right), we compare the ICM X-ray emission with the 1280 MHz radio emission. The radio halo broadly follows the X-ray surface brightness distribution, with both showing elongation along the NW–SE axis. The X-ray emission extends slightly further than the radio halo, while the radio and X-ray peaks coincide with an offset of less than 2$''$. \citet{pandge19} reported that the 235 MHz GMRT emission was confined within the SE cold front, similar to minihalos; however, our data reveal significant emission beyond the cold front at all frequencies. A notable similarity between the radio and X-ray distributions is a sloshing-like feature connected to the bright core. The GGM-filtered maps (Appendix~\ref{ggm_image}) further highlight an identical sharp edge in both the radio and X-ray maps.

\subsection{Thermodynamics maps}

We used the Adaptive Circular Binning (ACB) algorithm to map the projected gas properties of the ICM with a 10$''$ pixel scale. Circular regions were defined to include 200 net counts, leading to an extraction size of 10$''$ near galaxy centres to $\sim$33$''$ at the edges. Spectra were fitted with an absorbed APEC model, fixing $N_{\rm H}$ to the Galactic value and abundance to 0.3 Z$_\odot$. Foreground point sources were masked, except for the central galaxy. Details of the procedure for making the temperature maps are followed as described in \citet{Bott_24}.

The projected temperature map (Figure~\ref{xray_maps}, lower panel) shows strong variations, with an average temperature value of $\sim$ 11 keV but a distinctly non-isothermal distribution. Hot regions ($\sim$12 keV) extend along the SE–NW merger axis, while cooler patches ($\sim$8 keV) and a central drop to $\sim$5–6 keV mark the cold front (magenta arrow). These values agree with \citet{pandge19}, who reported upstream (11.65$\pm$2.0 keV) and downstream (6.0$\pm$1.0 keV) temperatures, with respect to cluster centre, and are consistent with \textit{Chandra} maps from \citet{Botteon2018}. Uncertainties are small ($\sim$0.2 keV) near the core, increasing to $\sim$1.9 keV at the outskirts. The radio halo correlates with hotter regions, though its peak overlaps with cooler central gas; a high-temperature streak along NE–SE coincides with the brightest halo emission, pointing to merger-driven activity. The entropy map shows a low-entropy core aligned with the central radio contours, suggesting dense, relatively cool gas surviving from a pre-merger cool core. Entropy rises asymmetrically to the NW and SE, consistent with merger heating, and the large-scale radio emission is cospatial with high-entropy regions.

\section{Spatial radio - X-ray correlation} \label{rad-xray-corr}

\begin{figure*}
    \includegraphics[width=0.52\textwidth]{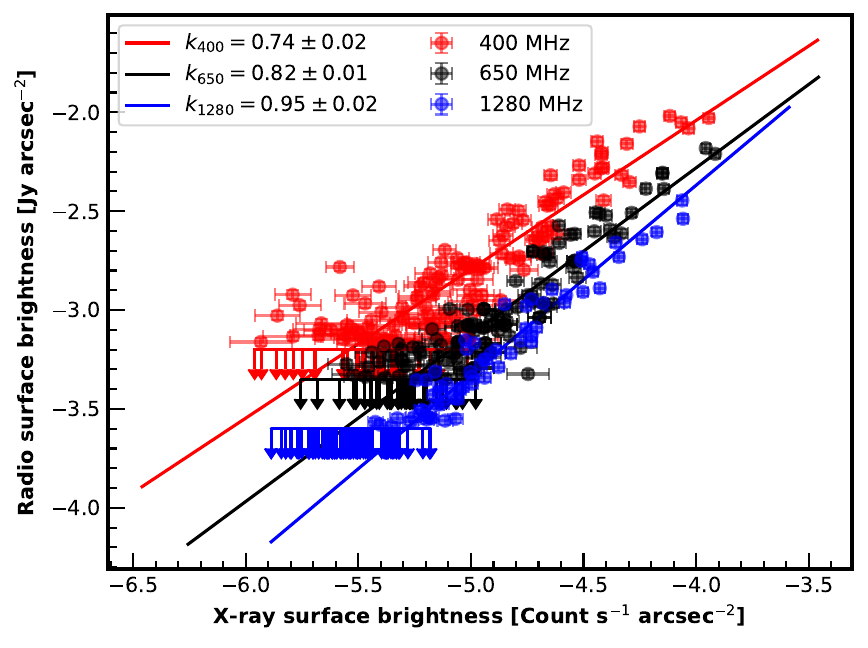}
    \includegraphics[width=0.49\textwidth]{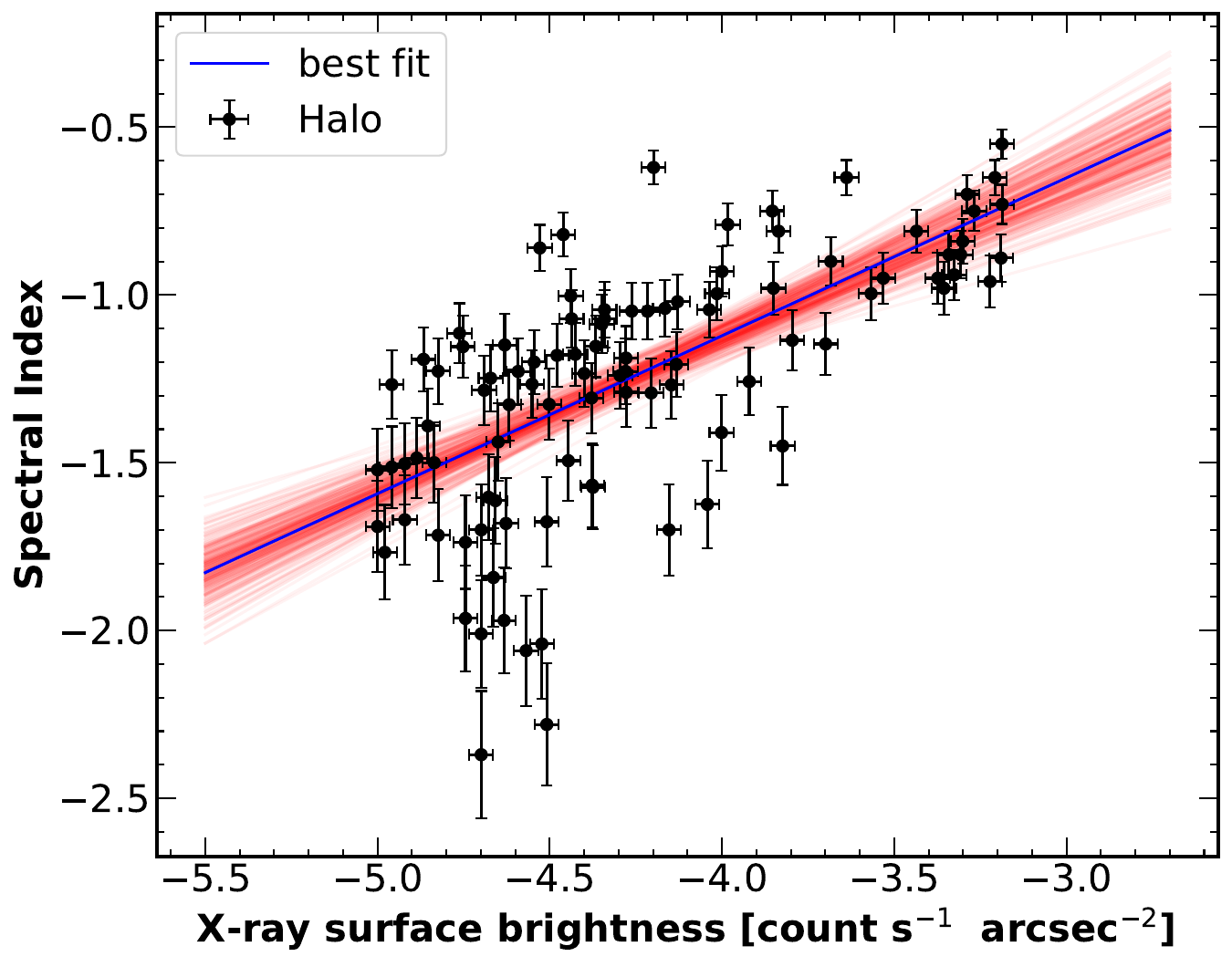}
     
    \caption{\textit{Left:} The radio vs. X-ray surface brightness correlation is shown for three different frequencies in three colours. The square boxes indicate the detection of the radio halo above 3$\sigma_{rms}$, and the arrows are the upper limits, \revone{with the same colour scheme as the corresponding observing frequencies}. The solid lines are the best-fit lines to the data at each frequency. \textit{Right:}The X-ray surface brightness and the spectral index correlation for the radio halo are shown. The blue solid line indicates the best fit for the regression, and the red regions are the posterior distribution from 1000 MCMC samples.}
    \label{ptp_corr_plot}
\end{figure*}

\begin{table*}
\centering
\caption{Best-fit slopes and Spearman (r$_{s}$) and Pearson (r$_{p}$) correlation coefficients of the radio and X-ray surface brightness data are summarised.}
 \begin{tabular}{@{}cccccccccc@{}}
 \hline\hline
 \multicolumn{0}{c}{}&\multicolumn{0}{c}{Frequency(MHz)}& \multicolumn{4}{c|}{3$\sigma$}& \multicolumn{4}{c|}{2$\sigma$} \\

\cline{3-10}
 & &slope (b) &$\sigma_{\rm int}$ & r$_{s}$ & r$_{p}$  & slope (b) &$\sigma_{\rm int}$ & r$_{s}$ & r$_{p}$ \\
\hline

 Halo & 400 & 0.72 $\pm$ 0.02 & 0.012 & 0.83 & 0.88 & 0.74 $\pm$ 0.02 & 0.012 & 0.83 & 0.88  \\

  &650 & 0.79 $\pm$ 0.03 & 0.010 & 0.83 & 0.89 & 0.82 $\pm$ 0.01 & 0.010 & 0.84 & 0.88   \\

  &1280 & 0.93 $\pm$ 0.02 & 0.005 & 0.95 &  0.97 & 0.95 $\pm$ 0.02 & 0.005 & 0.95 & 0.97  \\
  
\hline

Central & 400 & 0.65 $\pm$ 0.05 & 0.008 & 0.90 & 0.91 & - & -& - & - \\

& 650 & 0.72 $\pm$ 0.05 & 0.005 & 0.91 & 0.91 & - & - & - & -  \\

&1280 & 0.81 $\pm$ 0.04 & 0.006 & 0.95 &  0.96 & - & - & - & -  \\

\hline

Outer & 400 & 0.77 $\pm$ 0.05 & 0.015 & 0.70 & 0.73&  0.79 $\pm$ 0.03 & 0.01 & 0.71 & 0.75 \\

& 650 & 0.81 $\pm$ 0.05 & 0.009 & 0.83 & 0.85 & 0.84 $\pm$ 0.02 & 0.009 &  0.82 & 0.84  \\

&1280 & 0.92 $\pm$ 0.03 & 0.005 & 0.88 &  0.89 & 0.95 $\pm$ 0.06 & 0.006 & 0.92 & 0.94  \\

\hline
\end{tabular}
\label{ptp_results_tab}
\end{table*}

Observations have revealed a correlation between diffuse radio emission and X-ray emission in the central regions of galaxy clusters, offering insights into the interplay between the non-thermal components and the thermal ICM gas \citep[e.g.,][]{2001A&A...369..441G}. The spatially resolved correlation analysis has been a key tool to quantify the strength of the correlations, and has been studied in many radio halos \citep[e.g.,][]{2001A&A...369..441G, giacintucci05, 2022ApJ...933..218B, 2024arXiv240218654B, riseley22, santra24b, kale25, santra2026, Srikanth2026, biswas2026}. The relationship is generally described in the form of a power law:

\begin{equation}\label{ptp_eq}
    \log(\rm I_{\rm R}) = \rm a + b\log(I_{X})
\end{equation}

where b is the slope of the correlation and b \textless 1 implies that the gas density and temperature decrease more rapidly than the non-thermal components, whereas b \textgreater 1 suggests vice versa. We aim to determine the strength of this correlation, which reflects the mutual evolution between the thermal and the non-thermal components, and how the correlation slope varies from the cluster core to its outskirts.

We have made use of the publicly available software \texttt{PT-REX\footnote{\url{https://github.com/AIgnesti/PT-REX}}} (\texttt{Point-to-point TRend EXtractor}) \citep{2020A&A...640A..37I, ignesti22} to obtain the averaged radio (\revone{from IMG2, IMG5, and IMG8}) and X-ray surface brightness from a squared grid (15$''$), sampling the full extent of the diffuse emission. We retained only those cells where the radio and X-ray surface brightness is above the 3$\sigma_{\rm rms}$ level, and emission between 2$-$3$\sigma_{\rm rms}$ is included as upper limits. We fit equation~\ref{ptp_eq} to the extracted radio and X-ray SB values, using \texttt{LinMix} \citep{2007ApJ...665.1489K}, taking into account the uncertainties of both the dependent and independent variables, and upper limits on the dependent variable. The correlation strength is measured using the \texttt{Pearson} (r$_{\rm p}$) and \texttt{Spearman} (r$_{\rm s}$) correlation coefficients.

The radio and X-ray surface brightness show a positive correlation at all frequencies, with slopes b$_{400} = 0.74 \pm 0.02$, b$_{650} = 0.82 \pm 0.01$, and b$_{1280} = 0.95 \pm 0.02$, and a correlation coefficient of 0.89 (Figure~\ref{ptp_corr_plot}). Table~\ref{ptp_results_tab} lists the best-fit slopes and coefficients for all the frequencies. To examine spatial variation, we fitted equation~\ref{ptp_eq} separately for the inner and outer regions, divided at the e-folding radius (Section~\ref{SB_prof_halo}). The results (Table~\ref{ptp_results_tab}) show flatter slopes in the inner region than in the outer, independent of frequency. Similar radial trends have been reported for other merging clusters with giant radio halos \citep[e.g.,][]{2022ApJ...933..218B, bruno23, 2024arXiv240218654B, santra2026}.

The correlation slope has been studied from \delone{ultra-low} \revone{low-frequencies} ($\sim$144 MHz) to high frequencies ($\sim$3 GHz), showing mixed trends: some halos steepen from sublinear to linear with frequency \citep[e.g.,][]{rajpurohit21a,digennaro23}, while others remain constant \citep[e.g.,][]{hoang19}. For MACSJ0417, we find a steepening of the slope from 400 MHz (0.74) to 1280 MHz (0.95). This may partly reflect different sampling regions, since the halo is more extended at low frequencies and contracts at high frequencies. To test this, we redo the correlation analysis using only regions detected above 3$\sigma$ at all frequencies, yielding b$_{400} = 0.69 \pm 0.05$, b$_{650} = 0.78 \pm 0.03$, and b$_{1280} = 0.94 \pm 0.01$. These slopes are flatter than those in the Table.~\ref{ptp_results_tab}, but the overall trend remains the same.

We also examined the spatial correlation between the spectral index and the X-ray surface brightness, using similar regions. The relation is described as:

\begin{equation}
    \alpha = a' + b_{\alpha}\log(\rm I_{X})
\end{equation}

where a$'$ and b$_{\alpha}$ are the free parameters. We performed a linear fit in the halo region above the 3$\sigma$ level using \texttt{LinMix}. The results indicate an anti-correlation (r$_{s} = -0.67$) between the spectral index and X-ray surface brightness, with a slope of $-$0.55, consistent with trends reported for other giant radio halos (Figure.~\ref{ptp_corr_plot}, right panel). In contrast, some clusters show either a positive correlation \citep[e.g.,][]{botteon20a, rajpurohit21a} or no clear relation \citep[e.g.,][]{hoang21}. We find no evidence of deviation from a single trend, consistent with the spectral steepening seen in the spatially resolved spectral index map. At the outskirts, however, faint radio emissions may produce artificially steep indices, leading to an anti-correlation.

\section{Discussion}

\subsection{Origin of the radio halo}

The connection between the radio halo and cluster mergers is well established \citep[e.g.,][]{cassano10, kale13, kale15}; however, many aspects of the radio halo characteristics remain unclear. In the homogeneous re-acceleration model, the spectrum steepens with distance \citep{brunetti01}, and the steepening frequency can be expressed as:

\begin{equation}
    \nu_{s} \propto \tau_{acc}^{-2} \frac{\rm B}{(\rm B^{2} + \rm B_{\rm \rm IC}^{2})^{2}},
\end{equation}
    
with $\tau_{acc}$ is the acceleration timescale \citep{cassano05,brunetti11}, and B$_{\rm IC}$ is the CMB equivalent magnetic field. \revone{For a fixed acceleration timescale, the steepening frequency reaches a maximum at the critical (or minimum-loss) field $B_* = B_{\rm IC}/\sqrt{3}$, corresponding to the maximum radiative lifetime of relativistic electrons. Consequently, both larger and smaller magnetic-field strengths result in lower steepening frequencies and hence steeper observed spectra.}

The observed radial spectral steepening in MACSJ0417 likely reflects a combination of magnetic-field variations, spatial changes in the acceleration efficiency, and the energy-dependent distribution of cosmic-ray electrons. The fact that the steepening occurs at different radii in the 400--650 MHz and 650--1280 MHz spectral index maps suggests that a simple model with a spatially constant acceleration timescale may be insufficient. In particular, higher-frequency spectral pairs probe higher-energy electrons, which are more strongly affected by radiative losses. therefore exhibit spectral steepening closer to the cluster centre. The larger e-folding radius observed at lower frequencies is consistent with this picture, indicating that lower-energy electrons remain detectable out to larger cluster-centric distances. A detailed modelling of these effects is beyond the scope of this work.

\citet{2019JApA...40...17S} found that the integrated halo spectrum follows a single power law up to 18 GHz with $\alpha=-1.5$ without spectral curvature. Our deeper observations reveal a slightly flatter slope ($\alpha=-1.3$) and still no curvature. While the homogeneous re-acceleration model predicts a spectral steepening near a frequency, which is a few times the electron critical frequency \citep[e.g.,][]{cassano06, cassano10}, variations in $\tau_{acc}$ or $B$ along the line of sight can smooth the steepening. \revone{Combining the ATCA measurements \citep{2019JApA...40...17S} with our low-frequency data suggests a possible spectral steepening between 2--4 GHz. However, this inference should be treated with caution. The ATCA diffuse-emission flux densities were derived by measuring the total flux density in the central cluster region and subtracting the contributions from compact sources, rather than by integrating the halo emission within the same region adopted in this work. Furthermore, the high-frequency observations may be affected by residual missing-flux effects owing to the limited short-baseline coverage, potentially leading to an underestimate of the diffuse halo emission. Therefore, the location of the spectral steepening frequency remains uncertain, and the following estimates should be regarded as first-order constraints.} Assuming that the observed steepening reflects the intrinsic synchrotron spectrum of the halo, the maximum acceleration timescale can be estimated using the formalism of \citet{brunetti16}:

\begin{equation}
    \frac{\nu_{s}}{\rm GHz} \sim \left(\frac{420}{\tau_{acc}/\rm Myr}\right)^{2} (1 + z)^{-7},
\end{equation}

where $\nu_s$ is the spectral steepening frequency. A steepening frequency in the range 2--4 GHz implies an acceleration timescale of $\tau_{acc} \sim 80$--60 Myr. This is approximately five times shorter than that inferred for the Coma cluster, suggesting that particle acceleration in MACSJ0417 may be more efficient in order to compensate for the stronger inverse-Compton losses at higher redshift. A shorter acceleration timescale can be achieved if the particle mean free path (mfp) is substantially reduced relative to the Coulomb mean free path, thereby increasing the acceleration efficiency \citep{2014Natur.515...85Z,digennaro21}. Assuming non-resonant turbulent acceleration \citep{brunetti16}, the acceleration timescale can be expressed as

\begin{equation}
    \tau_{acc} \sim 450 \left(\frac{\psi}{0.5}\right)^{0.3} {\rm ~Myr},
\end{equation}

where $\psi = l_{\rm mfp}/l_{\rm A}$, and $l_{\rm A}$ and $l_{\rm mfp}$ denote the Alfv\'en scale and particle mean free path, respectively. Using the above estimate of $\tau_{acc}$ yields $\psi \lesssim 0.1$, significantly smaller than the reference value of $\sim0.5$ adopted in several previous studies \citep[e.g.,][]{brunetti20,nishiwaki24}. Similar constraints have recently been suggested for the high-redshift cluster El Gordo ($z=0.87$; \citealt{kale25}). Although tentative, MACSJ0417 ($z=0.445$) provides an important laboratory for testing particle acceleration models in the high-redshift Universe and may indicate higher acceleration efficiencies than predicted by standard scenarios.

The radio and X-ray emissions remain strongly correlated across frequencies, indicating that the non-thermal component declines more slowly than the thermal gas. The slope evolves from sublinear at low to nearly linear at high frequencies, supporting the change in e-folding radius and the frequency-dependent radial decline of radio emission. The observed radial steepening of the radio and X-ray correlation, with a flatter slope in the cluster central part and a steeper slope in the outer regions, is consistent with theoretical expectations from turbulent re-acceleration models. Under the assumption that the magnetic field scales with the thermal gas density, and accounting for the radial variation of turbulent energy flux, cosmic-ray content, and the relative contribution of inverse Compton losses, such models predict a steepening of the radio–X-ray correlation with increasing cluster radius \citep[e.g.,][]{2022ApJ...933..218B, 2024arXiv240218654B, santra2026}. This is driven by a combination of declining magnetic field strength and decreasing turbulent Mach number in the cluster outskirts. The radial trend we observe in MACSJ0417 is in qualitative agreement with these predictions. In merging clusters, turbulence-driven advection concentrates cosmic rays in the core, producing a central peak that declines radially \citep[e.g.,][]{ensslin11}. Variations in B(r) further modulate this trend, but their combined effect remains uncertain. Overall, the spectral and correlation results support turbulent re-acceleration, with evidence for inhomogeneous efficiency and radially varying non-thermal energy density.

\subsection{Constraining the pure hadronic scenario}

\begin{figure}
	\includegraphics[width=\columnwidth]{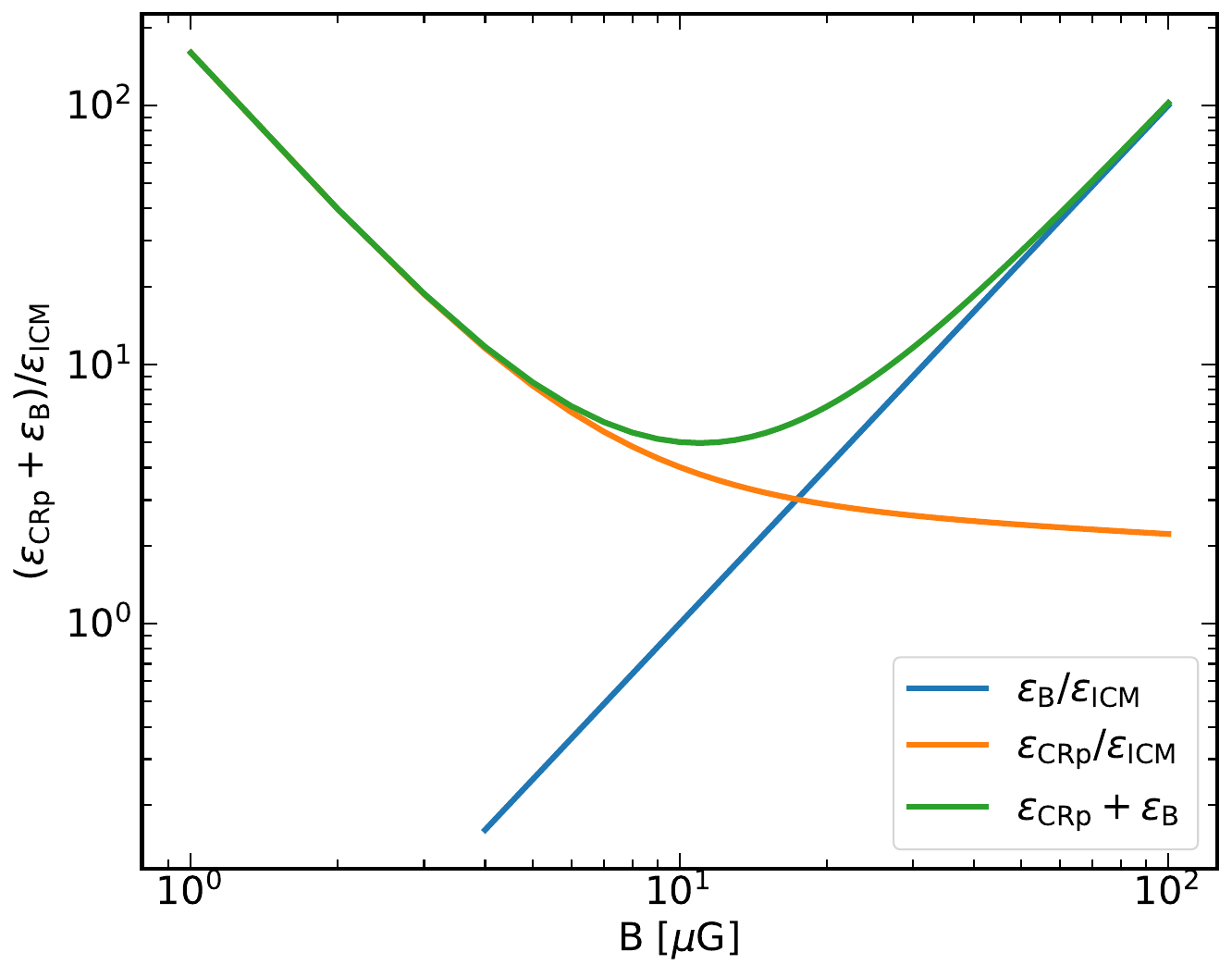}
        \caption{Energy densities of CRp and the magnetic field, required in a pure hadronic model without re-acceleration, plotted as a function of the magnetic field strength. The blue and orange curves represent the $\epsilon_{B}$ and $\epsilon_{CR}$, respectively, while the green line shows the sum of the two. All the energy densities are normalised to the ICM energy density $\epsilon_{ICM} = 4.6 \times 10^{-12}$ erg~cm$^{-3}$.}
    \label{macsj0417_had_model}
\end{figure}

In addition to turbulent re-acceleration, secondary electrons from pion decay between thermal protons and CRp (\textit{pp}) collision may contribute to diffuse emission in clusters \citep[e.g.,][]{dennison80, blasi99,keshet10,ensslin11}. Gamma-ray limits from the ICM \citep[e.g.,][]{jeltema09, ackermann2010, ackermann2012, ackermann2014} suggest that the direct \textit{pp}-induced emission is generally sub-dominant \citep{brunetti17, adam21}. However, secondary electrons can still seed the particle population for re-acceleration via turbulence \citep[e.g.,][]{brunetti17, pinzke17, nishiwaki22}. The moderately high-redshift cluster MACSJ0417, with its spectral properties and thermal to non-thermal correlation, challenges the purely hadronic origin and offers a direct test of hadronic models without gamma-ray constraints. Detailed theoretical calculations and background assumptions are described in the Appendix~\ref{CRps_calc}.

Figure~\ref{macsj0417_had_model} shows the ratio of CRp energy density ($\epsilon_{\rm CRp}$) to the ICM thermal energy density ($\epsilon_{\rm th}$) for the pure hadronic model, as a function of magnetic field. We also show the ratio of total non-thermal ($\epsilon_{\rm CRp} + \epsilon_{\rm B}$) to thermal energy density. For MACSJ0417, the average magnetic field B $<$ 10$\mu$G requires an unrealistic $\epsilon_{\rm CRp}/\epsilon_{\rm th} > 1$. Although $\epsilon_{\rm CRp}$ decreases with an increase in B, the minimum non-thermal energy density is still $\sim$7 times the ICM value, occurring at $B = 10\mu$G. The CRp energy budget implied here is far larger (60$–$70\%) than the few percent constrained by Fermi for Coma and other nearby clusters. This estimate does not include the re-acceleration of secondary electrons. Assuming spherical symmetry for both the ICM and CRp, the key parameter is the minimum CRp energy. With a conservative $E_{\rm min}=300$ MeV (threshold for CR$_{p–p}$ collisions), the CRp energy density is only $\sim$35\% lower than in Figure~\ref{macsj0417_had_model}. Extending the spectrum to $E_{\rm min}=1$ MeV and 20 keV boosts $\epsilon_{\rm CRp}$ by factors of 2.3 and 3.4, respectively. Thus, a purely hadronic origin is excluded: with $\epsilon_{\rm CRp}/\epsilon_{\rm th}$ only a few percent, secondary electrons contribute $<10\%$ of the halo emission. Nonetheless, hadronic collisions may still provide seed particles for turbulent re-acceleration \citep{brunetti11, pinzke17}.

\begin{figure*}
    \centering
    \includegraphics[width = 0.85 \textwidth]{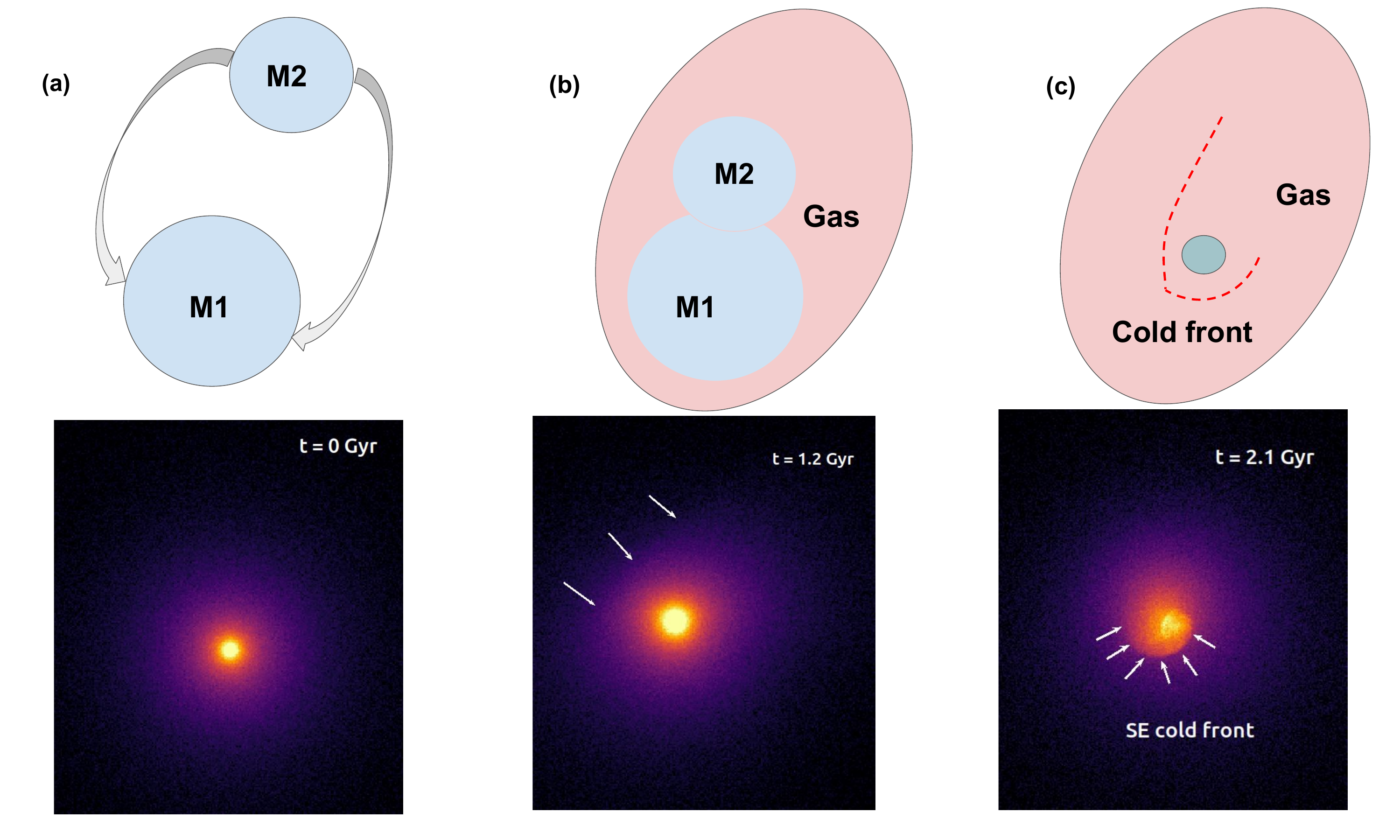}
    \caption{\textit{Top panels:} Schematic representation of the proposed merger scenario for MACSJ0417. (a) Two sub-clusters undergo a slightly off-axis collision nearly in the plane of the sky. (b) As clusters approach, the gas in between is heated, and they collide off-axis. (c) In the current configuration of MACSJ0417, the low mass sub-cluster may slosh around the cluster's main potential, creating the SE cold front, and also the main sub-cluster has retained all the gas. \textit{Bottom panels:} X-ray surface brightness slices (for the z projection) from an idealised simulation by \citet{zuhone18} of a 6:1 merger with an impact parameter of 500 kpc, taken at different times and used to support our sketched merger scenario of MACSJ0417.}
    \label{macsj0417_merger}
\end{figure*}

\subsection{Characterization of the R1 and R2}

R1 and R2 are detected at $\sim$2.9 Mpc from the cluster centre, with no optical counterparts, and a steep spectrum $\alpha \simeq -1.6$, consistent with the radio relic properties. \textit{XMM-Newton} data are too shallow to reveal shocks or density discontinuities, but their peripheral location and spectrum are consistent with relics. 
In the \revone{DSA} framework, the spectral index implies $\mathcal{M}$ = \revone{2.08 $\pm$ 0.08} (in Section~\ref{sec:int_spec}), unusual at such large radii where the gas density is very low. An accretion relic origin is unlikely, as simulations predict their Mach number to be $\mathcal{M} \gtrsim 10$, with a flatter injection spectrum ($\alpha \sim -0.5$). The observed steeper index instead suggests ageing or inefficient acceleration. Their radio power ($P_{1.4} \sim 10^{23}$ W Hz$^{-1}$) also exceeds expectations from accretion shocks ($\sim 10^{22}$ W Hz$^{-1}$; \citealt{hoeft07}), perhaps due to stronger fields/ projection effects.

The fraction of shock kinetic energy ($\xi_{e}$) channelled into the particle acceleration from the thermal pool is expressed as (\citealt{rajpurohit25a}):

\begin{eqnarray}
\nonumber
\xi_e = \frac{1}{C} 
\left( \frac{P}{\mathrm{W\ Hz^{-1}}} \right)
\left( \frac{A}{\mathrm{Mpc}^3} \right)^{-1}
\left( \frac{n_{e,d}}{10^{-4}\ \mathrm{cm}^{-3}} \right)^{-1} \\
\nonumber
\left( \frac{T_d}{7\ \mathrm{keV}} \right)^{-3/2}
\left( \frac{\nu}{1.4\ \mathrm{GHz}} \right)^{-\alpha}
\left( \frac{B}{\mu\mathrm{G}} \right)^{1+\alpha} \\
\left( \frac{B_{\mathrm{CMB}}^2}{B^2} + 1 \right)
\Psi(\mathcal{M}, T_d)
\end{eqnarray}

Here $C = 1.27 \times 10^{27}$ \citep{rajpurohit21a}, $P$ is the 1.4 GHz radio power, $A$ the emission area ($\pi/4 \cdot$ LLS$^{2}$; \citealt{hoeft07}), $n_{e,d}$ and $T_{d}$ the downstream density and temperature, and $\Psi(\mathcal{M},T_{d})$ a function of $\mathcal{M}$. At 1.4 GHz we derived $P_{1.4}$, $A$, and $\alpha$, adopting $n_{e,d}=10^{-4}$ cm$^{-3}$ and $T_{d}=1.7$ keV (using $T=1.3\langle T\rangle (1+1.5\frac{r}{r_{vir}})^{-1.6}$; \citealt{loken02}). For $\mathcal{M}\sim2.64$ and $B=0.1,\mu$G, the required electron acceleration efficiency is $\sim10^{-1}$, far above the values generally expected for weak merger shocks. Theoretical studies of DSA at low-Mach-number cluster shocks show that injection from the thermal pool is highly inefficient for $\mathcal{M}\lesssim3$, motivating scenarios in which relic emission is produced by the re-acceleration of pre-existing fossil electrons rather than by direct acceleration from the thermal pool \citep[e.g.,][]{kang11, kang_12, Pinzke2013}. Observational estimates of radio relic energetics also show that standard DSA from the thermal pool often requires unrealistically large efficiencies, approaching $\sim10$ per cent in several systems \citep[e.g.,][]{Vazza2014, 2020A&A...634A..64B}. Thus, the high radio luminosity inferred here remains difficult to explain within a simple thermal-pool DSA framework.

An alternative explanation is fossil radio plasma from a past AGN activity, consistent with the absence of an optical counterpart and the steep spectra expected from aged electrons. Confirmation would require evidence of (i) a flatter-spectrum injection edge ($\alpha \sim -0.6$ to $-0.8$), (ii) strong spectral curvature, or (iii) polarisation observation. Deeper low-frequency and sensitive X-ray observations will be essential to establish the nature of this feature.

\subsection{Merger history and radio emission}

The dynamical state of MACSJ0417 provides important clues to its diffuse radio emission. \citet{2012MNRAS.420.2120M} found the X-ray morphology consistent with a high-velocity encounter, akin to the Bullet cluster \citep[e.g.,][]{markevitch01}. \textit{Chandra} and GMRT data \citep{pandge19} revealed cold fronts and diffuse emission, suggesting a dissociative merger where the subcluster lost its gas while the main cluster retained it. \citet{2017MNRAS.464.2752P} derived morphological parameters placing the system between relaxed and merging, with an elongated SE–NW X-ray structure. \citet{giacintucci17} reported central entropy $K_{0}=30$ keV~cm$^{2/3}$~arcsec$^{-2}$, placing the cluster marginally in the cool-core regime. 

We propose that MACSJ0417 is a post-merger system where a subcluster from the NW merged with the main cluster along the NW–SE axis (schematic in Figure~\ref{macsj0417_merger}). To interpret our results, we use the Galaxy Cluster Merger Catalog \citep{zuhone18}, which provides binary merger simulations over broad parameter ranges. The bottom panels of Figure~\ref{macsj0417_merger} show X-ray slices from a 6:1 merger with a 500 kpc impact parameter, reproducing features similar to MACSJ0417. The elongated SE–NW morphology, surface brightness discontinuities, and hot outskirts suggest that a less massive subcluster (1:6) approached from the NW, underwent strong ram-pressure stripping, and left a disturbed trail while its dark matter halo advanced inward. This encounter likely generated turbulence in the outer ICM, forming the giant radio halo, while gravitational perturbations displaced the main core, inducing sloshing motions that produced the SE cold front. The NW discontinuity may trace either the mirror of this sloshing spiral or the remnant contact discontinuity of the stripped core. The deep potential of the primary cluster preserved the $\sim$6–7 keV cool core despite the merger. Simulations show cool cores can survive unequal-mass, off-axis mergers with high angular momentum \citep{valdarnini21}, and recent observations confirm cool cores in disturbed systems \citep{Somboonpanyakul_2021}. Within the evolutionary framework of \citet{molendi23}, cool cores and non-cool cores are better understood as dynamical states between which clusters can move, rather than distinct subclasses. In this picture, MACSJ0417 may be viewed as a cool-core remnant where the low-entropy, metal-enriched core has survived the merger activity, only partially disrupted, while the outer ICM has been sufficiently stirred to sustain giant radio halo emission. This interpretation finds observational support in \citet{gitti25}, where PLCKG287 has been identified as a former heated cool core, in which the central metallicity peak persists as a fossil of the cool core enrichment of the past despite significant shock heating. Together, these studies suggest that the outcome of a merger depends sensitively on the mass ratio and impact parameter, and that MACSJ0417 may represent an earlier stage along this evolutionary sequence, one where the cool core is being progressively disrupted but has not yet transitioned to a fully non-cool-core state.

\section{Summary}\label{summary}

We present uGMRT and MeerKAT observations of the merging cluster MACSJ0417 across 300–1700 MHz. These sensitive observations, combined with \textit{XMM-Newton}, provide new insights into the thermal–non-thermal interplay in the ICM. The key results are summarised below:

\begin{enumerate}

\item We present the deep radio images (5$''$–30$''$) of MACSJ0417, revealing halo emission of $\sim 1.75$ Mpc at 400 MHz along the NW–SE axis. We also detect elongated emission at 2.9 Mpc from the centre, with no optical counterpart and a projected size of $\sim 1$ Mpc.

\item The radial surface brightness profile of the halo is well described by a single exponential profile, with a frequency-dependent e-folding radius. Our analysis also confirms the SE discontinuity $\sim$0.7$'$ from the centre, coincident with the X-ray edge, showing a factor of 4$-$5 drop in surface brightness within 100 kpc.

\item The integrated halo spectrum follows a single power law ($\alpha \simeq -1.3$) between 400--1280 MHz, and R1 and R2 have $\alpha \sim -1.6$. \revone{The spatially resolved map of the halo shows significant spectral index fluctuations and a radial steepening from the centre to the outskirts. While residual systematic effects cannot be entirely excluded, the observed scatter suggests inhomogeneous conditions within the synchrotron-emitting volume.}

\item The \textit{XMM-Newton} map shows an elongated SE–NW morphology over $\sim$1.9 Mpc with a bright comet-like core. The radio halo traces the thermal gas, confirming a strong morphological correlation, with the temperature map revealing high cluster temperatures ($\sim$11 keV).

\item A strong positive radio vs X-ray correlation is found, with the slope evolving from sublinear at low to linear at high frequencies. An anti-correlation between spectral index and X-ray brightness confirms radial spectral steepening.

\item \revone{The spectral and radio-to-X-ray correlation studies are broadly consistent with the turbulent re-acceleration scenario, although they also suggest variations in the physical conditions of the ICM. If the spectral steepening inferred from the available high-frequency measurements is intrinsic, the implied electron acceleration timescale is shorter than that observed in nearby radio halos, pointing toward a more efficient acceleration process at intermediate redshift.}

\item \revone{The origin of R1 and R2 remains uncertain. Their elongated morphology, large projected extent, and lack of optical counterparts are broadly consistent with a relic interpretation; however, the absence of polarisation information, spectral gradients, and direct shock constraints prevents a firm classification. Future deep radio and X-ray observations will be crucial for determining their nature.}

\item  Multi-wavelength data reveal a peculiar merger history, with a minor merger (6:1 mass ratio) likely preserving the core while producing the SE cold front. The presence of a giant radio halo with a cool core places MACSJ0417 in a rare regime.

\end{enumerate}


\begin{acknowledgments}

We thank the anonymous referee for their constructive comments that have improved the clarity of the paper. R.S. and R.K. acknowledge the support of the Department of Atomic Energy, Government of India, under project no. 12-R\&D-TFR-5.02-0700. R.S. also acknowledge the support of the Department of Atomic Energy, Government of India, under project no. RTI4019. R.K. also acknowledges the support from the SERB Women Excellence Award WEA/2021/000008. M.B. acknowledges support from the ERC CoG $\vec{B}$ELOVED, GA N.101169773. M.B. acknowledges the financial contribution from the INAF GO grant 1.05.24.02.10 {\it Extended Radio Emission in Galaxy Clusters at deep focus with MeerKAT}. We thank the staff of the GMRT that made these observations possible. The GMRT is run by the National Centre for Radio Astrophysics (NCRA) of the Tata Institute of Fundamental Research (TIFR). This research made use of the NASA/IPAC Extragalactic Database (NED), which is operated by the Jet Propulsion Laboratory, California Institute of Technology, under contract with the National Aeronautics and Space Administration. The MeerKAT telescope is operated by the South African Radio Astronomy Observatory, which is a facility of the National Research Foundation, an agency of the Department of Science and Innovation. Based on observations obtained with XMM-Newton, an ESA science mission with instruments and contributions directly funded by ESA Member States and NASA.

\end{acknowledgments}





%
\facilities{uGMRT, MeerKAT, XMM-Newton}

\software{Astropy \citep{2013A&A...558A..33A,2018AJ....156..123A,2022ApJ...935..167A}, ApLPy \citep{2012ascl.soft08017R}, CASA\citep{mcmullin07}, ds9 \citep{ds9_2000}, 
XMM-SAS \citep{xmmsas04}}


\appendix

\section{GGM image of radio and X-ray maps}\label{ggm_image}

We have shown the GGM-filtered image of both the X-ray and radio (650 MHz) maps.

\begin{figure*}
    \includegraphics[width=0.48\textwidth]{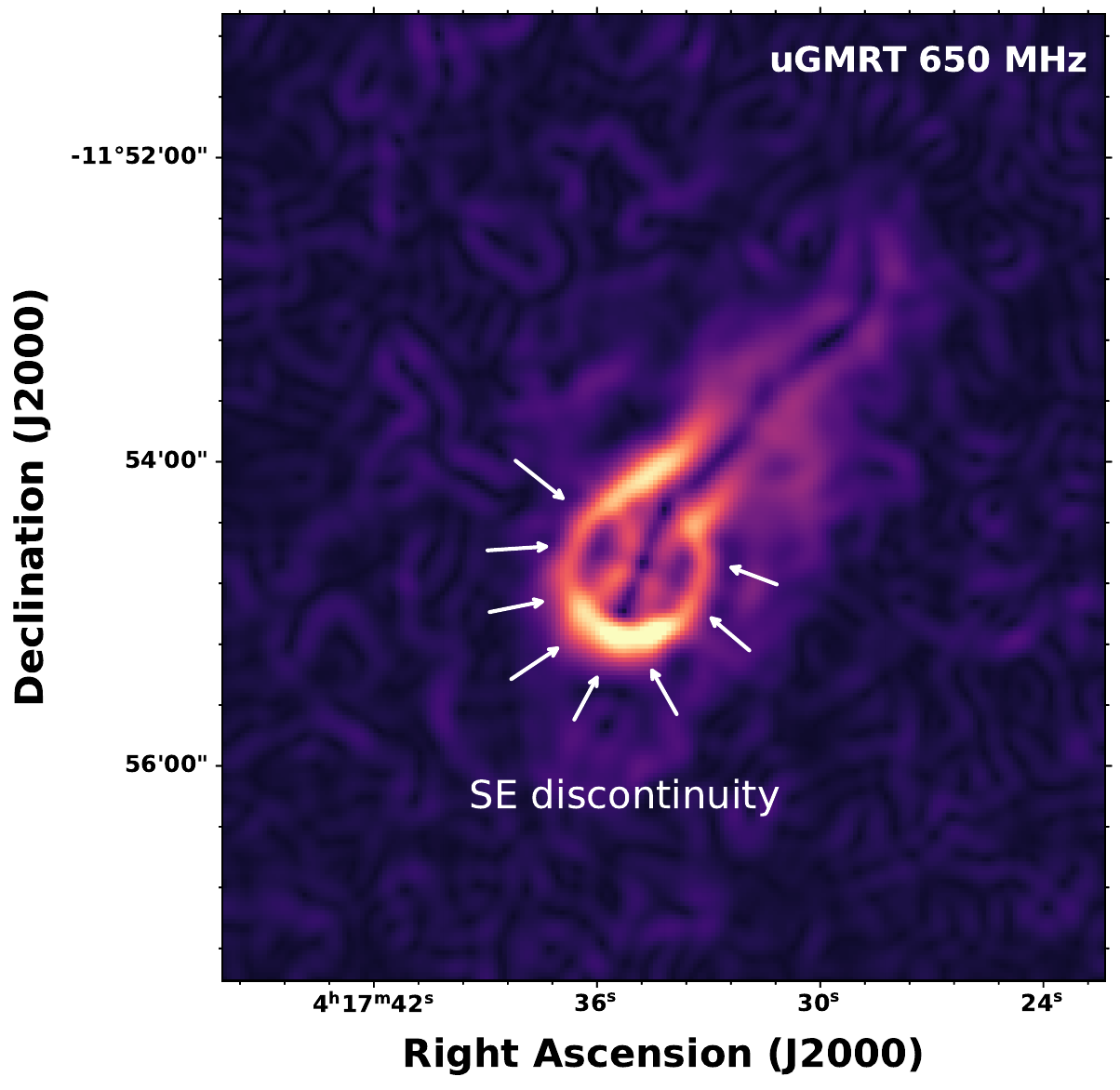}
    \includegraphics[width=0.39\textwidth]{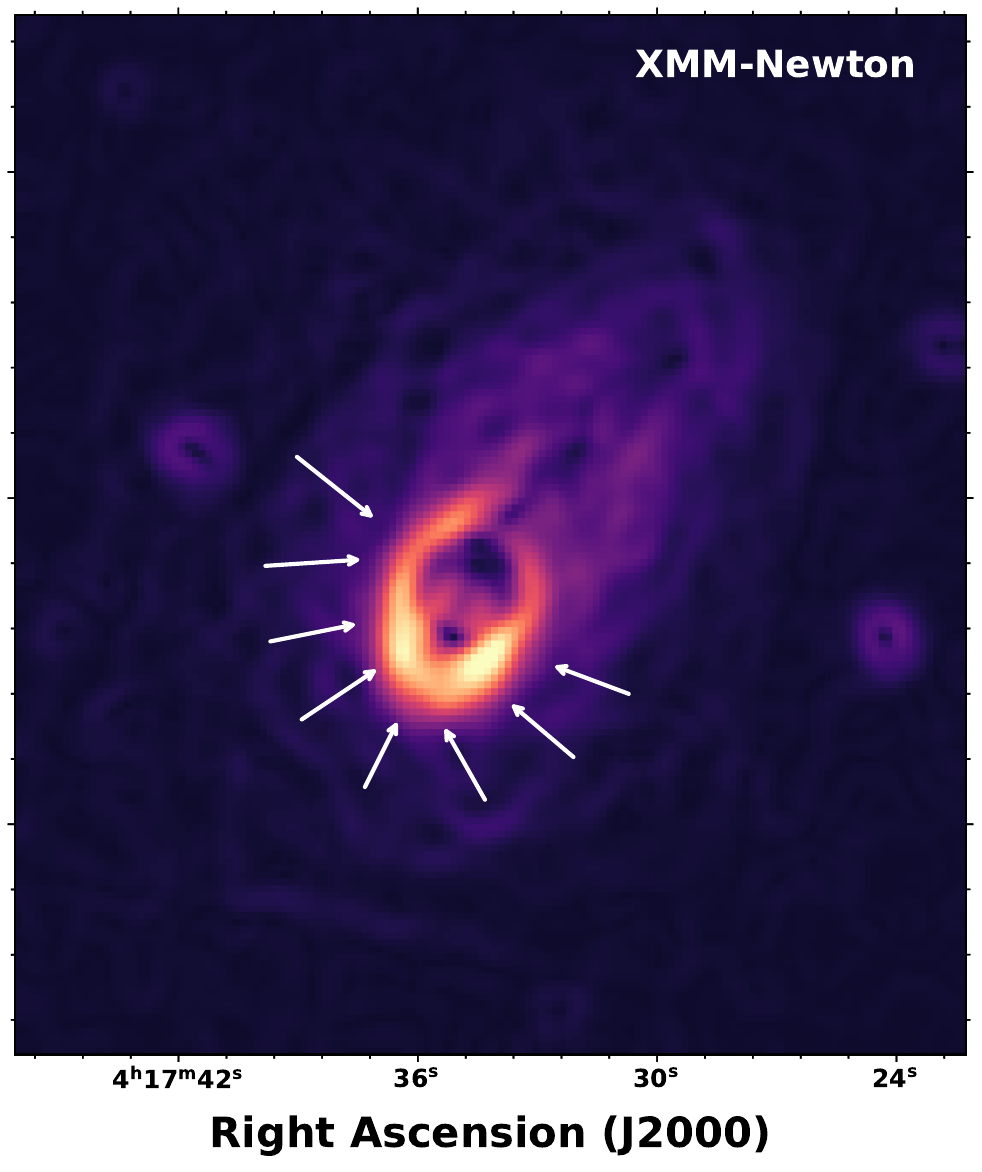}
     
    \caption{\textit{Left:} The \textit{ggm-}filtered image with $\sigma = $ 4 pixels on the uGMRT 650 MHz radio map is shown here. The central bright part is masked from the image domain. The white arrows highlight the location of the SE discontinuity.  \textit{Right:} The same is shown for the \textit{XMM-N}ewton image with white arrows highlighting the discontinuity.}
    \label{macsj0417_ggm}
\end{figure*}

\section{Spectral index error maps}\label{sec:spec_err_maps}

We show here the spectral index uncertainty maps, corresponding to Figure~\ref{spec_maps}

\begin{figure*}
    \includegraphics[height = 15cm, width = 0.85 \textwidth]{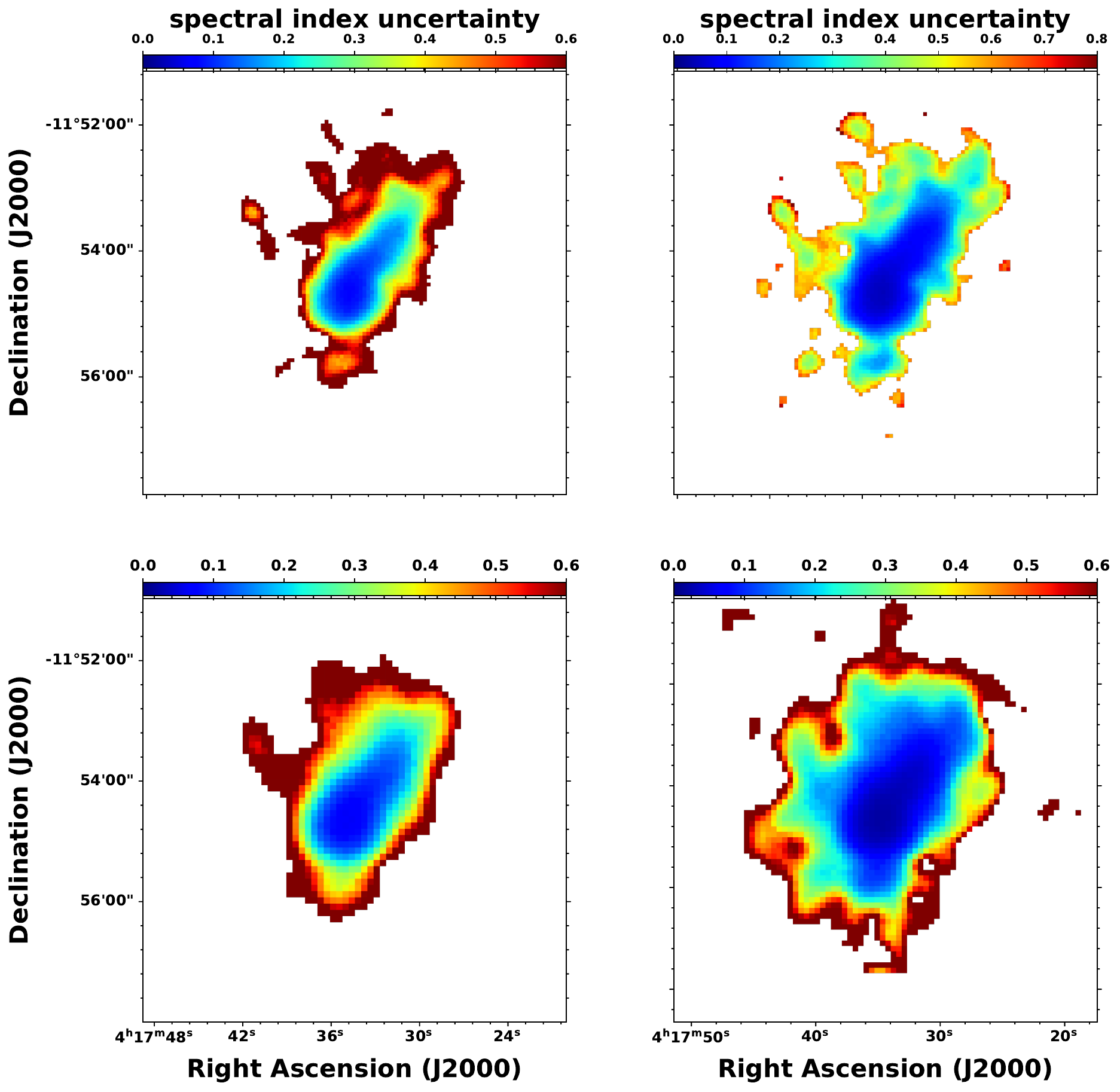}
     
   \caption{\textit{Upper left:} Spectral index error map between 400 and 650 MHz, at a resolution of 15$''$. \textit{Upper right:} The error map is shown, but at 15$''$ resolution between 650 and 1280 MHz. The errors mostly dominate below 0.1, in the central region. \textit{Lower left:} The spectral index uncertainty map between the uGMRT frequencies is shown at 30$''$ resolution. \textit{Lower right:} spectral index error map between 650 and 1280 MHz, at a resolution of 30$''$.}
    \label{spec_maps_err}
\end{figure*}

\section{Secondary electron contribution}\label{CRps_calc}

We aim to constrain the contribution from cosmic ray protons behind the observed non-thermal emission in a pure hadronic scenario, following \citet{2021A&A...650A..44B}. The stationary spectrum of the electrons and protons produced during the hadronic collision in the ICM is given by \citep{dolag00}:

\begin{equation}
N_e^{\pm}(p,t) = \frac{1}{\sum_{\mathrm{rad},i} \left| \frac{dp}{dt} \right|} \int_{p} Q_e^{\pm}(p',t) \, p' \, dp'
\end{equation}

where $p$ is the particle momentum, $\left| \frac{dp}{dt} \right|_{\mathrm{rad},i}$ are the radiative and Coulomb losses, and $Q_e^{\pm}(p,t)$ is the electrons and positron injection rate, which is defined as \citep[][and references therein]{brunetti14}:

\begin{eqnarray}
\nonumber
Q_e^{\pm}(p,t) = \frac{2.1712 \, m_{\pi}^2 \, n_{\mathrm{th}} \, c^2}{m_{\pi}^2 - m_{\mu}^2}
\int_{E_{\mathrm{min}}} \int_{p^*} \frac{dE_{\pi} \, dp}{E_{\pi} \, \bar{\beta}_{\mu}} \, \beta_p \, \\
\nonumber
N_p(p,t)
\times \frac{d\sigma_{\pm,0}}{dE}(E_{\pi}, E_p) \, \\
F_e(E_e, E_{\pi})
\end{eqnarray}

where $\pi$, $\mu$, and $p$ refer to pions, muons, and protons, respectively, $\bar{\beta}_{\mu} = \sqrt{1 - \left( \frac{m_{\mu}}{\bar{E}_{\mu}} \right)^2}$, $\bar{E}_{\mu} = \frac{E_{\pi}}{0.5428} \left[ 1 - \left( \frac{m_{\mu}}{m_{\pi}} \right)^2 \right]$, and $\frac{d\sigma_{\pm,0}}{dE}(E_{\pi}, E_p)$ is the differential inclusive cross-section for the production of $\pi^+$, $\pi^-$, and $\pi^0$ (assuming cross section provided in \citealt{brunetti17}), and $F_e(E_e, E_{\pi})$ can be defined as \citet{brunetti05} (their equation~36 and 37). Note that we have assumed the order-of-magnitude estimates of the electron/positron production rate using the cross section of the pp interaction (30 mb)
and the energy branching ratio of $\sim$0.1, consistent with the inelasticity and secondary production efficiencies derived by \cite{kelner06}. 

The distribution of the thermal protons can be derived from the X-ray observations, $n_{\mathrm{th}}(r)$, from the X-ray surface brightness profile $I_X(r)$ (index $\beta = 0.60$, core radius $r_c = 288$~kpc, and central proton density number $n_{\mathrm{th},0} = 2.9 \times 10^{-3}~\mathrm{cm}^{-3}$ \citep{giacintucci17}. We also assumed that the magnetic field scales with the thermal proton density as:

\begin{equation}\label{b-ne-eq}
B(r) = B_0 \left[ \frac{n_{\mathrm{th}}(r)}{n_{\mathrm{th},0}} \right]^{\eta}
\end{equation}

where $\eta = 0.5$ is fixed \citep{bonafede10}, although this simplistic assumption has recently been under scrutiny due to numerous observational evidence from thermal to non-thermal correlation studies \citep[e.g.,][]{balboni23}.

The secondary CRe injected in the ICM produces synchrotron radio emission with an emissivity given by:

\begin{align} 
\label{eq:sync_eq}
j_R(\nu, r) &= \frac{\sqrt{3} \, e^3}{m_e c^2}
\int_{0}^{\pi/2} \sin^2\theta \, d\theta
\int N_e^{\pm}(p) \, F\left( \frac{\nu}{\nu_c} \right) dp \nonumber \\
&\propto N_p(p,r) \, n_{\mathrm{th}}(r) \,
\frac{B(r)^{1+\alpha}}{B(r)^2 + B_{\mathrm{IC}}^2} \, \nu^{-\alpha} \nonumber \\
&\propto n_{\mathrm{th}}\,K_{p} \,
\frac{B^{1+\alpha}}{B^{2} + B_{\mathrm{CMB}}^{2}} \, \nu^{-\alpha}
\end{align}

where $K_{p}$ = normalization of CRp spectrum ($N_{p}(E) = K_{p} E^{-\alpha_{p}}$), $B_{\mathrm{CMB}} = 3.25 (1+z)^{2}~\mu\mathrm{G}$. From observed $I_{\nu}$ and path length $L$, $I_{\nu} \approx j_{\nu} L$. Inverting for $K_{p}$, the equation.~\ref{eq:sync_eq} becomes: 

\begin{equation} \label{eq:C5}
    K_{p} \propto \frac{I_{\nu}}{n_{\mathrm{th}} L}
    \frac{B^{2} + B_{\mathrm{CMB}}^{2}}{B^{1+\alpha} \, \nu^{-\alpha}}
\end{equation}

For a pure power-law proton spectrum, the total CRp energy content is then computed as:

\begin{equation} \label{eq:C6}
    \epsilon_{\mathrm{CRp}} \equiv \int_{E_{\min}}^{E_{\max}} N_{p}(E) \, E \, dE
    = K_{p} \,
    \frac{E_{\min}^{\, 2-q_{p}} - E_{\max}^{\, 2-q_{p}}}{q_{p} - 2}
\end{equation}

We adopted $p_{\mathrm{min}} = 1.0\,m_p c$, corresponds to a E$_{min} = 220$ Mev. Combining the equation.~\ref{eq:C5}, \ref{eq:C6}:

\begin{equation}
    \epsilon_{\mathrm{CRp}} \propto C_{CRp} \frac{B^{2} + B_{\mathrm{CMB}}^{2}}{B^{1+\alpha}}
\end{equation}

The magnetic energy is computed as $\epsilon_{\rm B} = \rm B^{2}/8\pi$, and each of the energy densities are normalized with $\varepsilon_{\mathrm{ICM}} = \frac{3}{2} n_{\mathrm{th}} k_{B}T$, the thermal energy of the ICM.


\bibliography{sample701}{}
\bibliographystyle{aasjournalv7}



\end{document}